\documentclass[11pt,a4paper]{article}
\usepackage{jheppub}
\usepackage{rotate}
\usepackage{lscape}
\usepackage{epsfig}
\usepackage{axodraw}
\usepackage{feynmf}
\usepackage{amsmath}
\usepackage{multirow}
\usepackage{subfiles}
\usepackage{subfig}

\newcommand{\bq}{\begin{equation}}
\newcommand{\eq}{\end{equation}}
\newcommand{\bqa}{\begin{eqnarray}}
\newcommand{\eqa}{\end{eqnarray}}
\newcommand{\baa}[1]{\begin{array}{#1}}
\newcommand{\eaa}{\end{array}}

\newcommand{\sss}[1]{\scriptscriptstyle{#1}}

\def\mtp{m_t}

\newcommand{\stw}{s_{\sss{W}}  }

\newcommand{\stws}{s^2_{\sss{W}}}
\newcommand{\stwsb}{\bar{s}^2_{\sss{W}}}

\newcommand{\ctws}{c^2_{\sss{W}}}
\newcommand{\ctwsb}{\bar{c}^2_{\sss{W}}}

\def\stw{s_{\sss W}}

\def\MSbar{$\overline{\mathrm{MS}}\ $}

\def\gsim{\mathrel{\raise.3ex\hbox{$>$\kern-.75em\lower1ex\hbox{$\sim$}}}}
\def\lsim{\mathrel{\raise.3ex\hbox{$<$\kern-.75em\lower1ex\hbox{$\sim$}}}}

\def\dd{{\mathrm d}}
\def\MSbar{$\overline{\mathrm{MS}}\ $}

\title{QCD parton showers and NLO EW corrections to Drell-Yan}

\author[a]{Peter~Richardson,}
\author[b]{Renat~R.~Sadykov,}
\author[b]{Andrey~A.~Sapronov,}
\author[c]{Michael~H.~Seymour,}
\author[d]{Peter~Z.~Skands,}

\affiliation[a]{Institute for Particle Physics Phenomenology, University of Durham, \\
South Rd, Durham DH1~3LE, United Kingdom}
\affiliation[b]{Joint Institute for Nuclear Research, \\
Joliot-Curie 6, Dubna, Moscow region, Russia}
\affiliation[c]{School of Physics and Astronomy, The University of Manchester, \\
Manchester, M13~9PL, United Kingdom}
\affiliation[d]{Theoretical Physics, CERN, \\
CH-1211, Geneva 23, Switzerland}

\emailAdd{peter.richardson@durham.ac.uk}
\emailAdd{renat.sadykov@cern.ch}
\emailAdd{andrey.sapronov@cern.ch}
\emailAdd{michael.seymour@manchester.ac.uk}
\emailAdd{peter.skands@cern.ch}

\abstract{We report on the implementation of an interface between the SANC generator
framework for Drell-Yan hard processes, which includes next-to-leading order
electroweak (NLO EW) corrections,  and the Herwig++ and Pythia8 QCD parton
shower Monte Carlos.  A special aspect of this implementation is that the
initial-state shower evolution in both shower generators has been augmented to
handle the case of an incoming photon-in-a-proton, diagrams for which appear at
the NLO EW level.  The difference between shower algorithms leads to residual
differences in the relative corrections of 2--3\% in the $p_T(\mu)$
distributions at $p_T(\mu)\gsim50$~GeV (where the NLO EW correction itself is of
order 10\%).}


\arxivnumber{1011.5444}

\notoc

\begin{document}

\maketitle
\flushbottom

\section{Introduction}

At high energy hadron colliders, studies of Drell--Yan (DY) like processes are
of great importance. They are crucial for the understanding of QCD and electroweak
(EW) \mbox{interactions} in hadron-hadron collisions.  Drell--Yan processes have large
cross sections and clean signatures in the detectors. They are used for
monitoring of the collider luminosity and calibration of detectors. DY is a
reference process for measurements of EW boson properties at hadron colliders.
Combination of accurate experimental measurements of these processes with
elaborated theoretical predictions allows the extraction of parton
density functions (PDFs) in the kinematical regions which were not yet accessed
in DIS experiments.  DY processes provide an important background to many other
processes studied at hadron colliders including searches for Higgs scalar as
well as for $W'$ and $Z'$ bosons in particular. All this gives a strong
motivation to have an advanced high precision theoretical description of DY.
The experimental precision of DY measurements at the LHC can reach the 1\% level.
That means that the accuracy of the theoretical predictions needs to be even
higher.  For this reason it is obvious that QCD and electroweak radiative
corrections should be taken into account.  

This article presents the results of application of parton shower algorithms to
a hard process that was calculated with electroweak radiative corrections in
the SANC system~\cite{Andonov:2004hi, Andonov:2008ga}.  The showering procedure
was applied to the Drell--Yan processes:
\begin{align}
\begin{split}
pp&\to(W^{+})\to l^{+}\nu_{l}(\gamma)+X, \\
pp&\to(\gamma,Z)\to l^+l^-(\gamma)+X,
\end{split}
\label{eqn:dy_process}
\end{align}
where $X$ represents hadrons and $l$ is one of $e, \mu, \tau$. The parton
shower algorithms implemented in the general-purpose Monte Carlo generators
Pythia8~\cite{Sjostrand:2007gs} and Herwig++~\cite{Bahr:2008tf} were used for
these processes. It is worth noting that these two programs use essentially
different parton shower algorithms: Pythia8 uses an evolution scheme based on
transverse-momentum ordering~\cite{Pythia8PHP:www} and Herwig++ uses the
coherent branching algorithm based on angular ordering of emissions in the
parton shower~\cite{Bahr:2008pv}.

Earlier the combination of the effects due to parton showers (PS) and due to EW
radiative corrections for charged current process was considered
in~\cite{Balossini:2008cs,Balossini:2009sa}.  In those studies an interface
between HORACE~\cite{CarloniCalame:2003ux,horace:www} and fortran event
generator Herwig~\cite{Corcella:2000bw} was developed.  Here, we present
studies in which the SANC generator~\cite{Arbuzov:2005dd, Arbuzov:2007db} is
used for the treatment of complete NLO EW corrections with interfacing it
to the Herwig++~v2.4.0 and Pythia8~v.130 generators to apply parton showers. 

The paper is organized as follows. In the next sections we describe the chain
of simulations and other topics associated with the calculation procedure.  In
section~\ref{sancmc_ssect} the relevant features of the SANC MC event generators
are described. Section~\ref{psmc} discusses aspects of the parton showers that
are added by Pythia8 or Herwig++. Numerical cross checks and results are given
in section~\ref{Scheme_sect}. In section~\ref{sum_sect} the obtained results and
prospects are discussed.

\section{The Drell-Yan processes in SANC  \label{sancmc_ssect}}

The SANC system~\cite{Arbuzov:2005dd,Arbuzov:2007db} provides tools for calculating 
the differential cross sections of the Drell-Yan
processes taking into account the complete (real and virtual)
${\mathcal{O}}(\alpha)$ electroweak radiative corrections. Here we give a
brief summary of the main properties of this framework. 

All calculations are performed within the OMS (on-mass-shell) renormalization
scheme \cite{Bardin:1999ak} in the $R_\xi$ gauge which allows an explicit control
of the gauge invariance by examining a cancellation of the gauge parameters in the
analytical expression of the squared matrix element.

We subdivide the total EW NLO cross section of Drell-Yan process at the
partonic level for observables $\vec{X}$ ($\vec{X} = (x_1,...,x_n)$ is a
generic observable which is a function of the final-state momenta) into
four terms:
\begin{align}
\frac{d^n\sigma^{\mathrm{NLO EW}}}{d\vec{X}} = \frac{d^n\sigma^{\mathrm{Born}}}{d\vec{X}}+
\frac{d^n\sigma^{\mathrm{virt}}}{d\vec{X}}\biggl(\lambda\biggr)
+\frac{d^n\sigma^{\mathrm{soft}}}{d\vec{X}}\biggl(\lambda,\bar{\omega}\biggr)
+\frac{d^n\sigma^{\mathrm{hard}}}{d\vec{X}}\biggl(\bar{\omega}\biggr),
\end{align}
where $\sigma^{\mathrm{Born}}$ is the Born level cross-section,
$\sigma^{\mathrm{virt}}$ is a contribution of virtual(loop) corrections,
$\sigma^{\mathrm{soft}}$ corresponds to a soft photon emission and
$\sigma^{\mathrm{hard}}$ is a contribution of a hard (real) photon emission. The
terms with auxiliary parameters $\bar{\omega}$ (photon energy which separates
phase spaces associated with the soft and hard photon emission) and
$\lambda$ (photon mass which regularizes infrared divergences) cancel out after
summation and the differential EW NLO cross-section for infrared-safe
observables does not depend on these parameters~\cite{Greco:1980mh,Bohm:1982hr,Denner:1991kt}.

\begin{figure}

\begin{minipage}[b]{0.5\linewidth}
\centering
\begin{fmffile}{tree_dy_nc}
  \begin{fmfgraph*}(40,25)
    \fmfleft{i2,i1}
    \fmfright{o1,o2}
    \fmflabel{$q$}{i1}
    \fmflabel{$\bar{q}$}{i2}
    \fmflabel{$\ell^+$}{o1}
    \fmflabel{$\ell^-$}{o2}

    \fmf{fermion}{i1,v1,i2}
    \fmf{fermion}{o1,v2,o2}
    \fmf{photon,label=$\gamma,,Z^0$}{v1,v2}
  \end{fmfgraph*}
\end{fmffile}
\end{minipage}
\hspace{-1.5cm}
\begin{minipage}[b]{0.5\linewidth}
\centering
\begin{fmffile}{tree_dy_cc}
  \begin{fmfgraph*}(40,25)
    \fmfleft{i2,i1}
    \fmfright{o1,o2}
    \fmflabel{$u$}{i1}
    \fmflabel{$\bar{d}$}{i2}
    \fmflabel{$\ell^+$}{o1}
    \fmflabel{$\nu_{\ell}$}{o2}

    \fmf{fermion}{i1,v1,i2}
    \fmf{fermion}{o1,v2,o2}
    \fmf{boson,label=$W^+$}{v1,v2}
  \end{fmfgraph*}
\end{fmffile}
\end{minipage}

\vspace*{1cm}
\caption{Feynman graphs for tree level Drell-Yan process, where \(u\) and \(d\) represent generic up and down type quarks respectively.}
\label{fig:fdiag_tree_dy}
\end{figure}
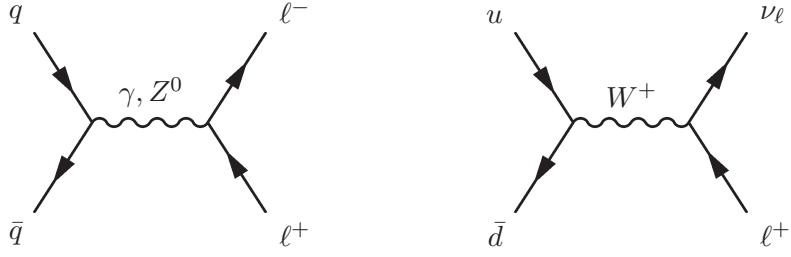

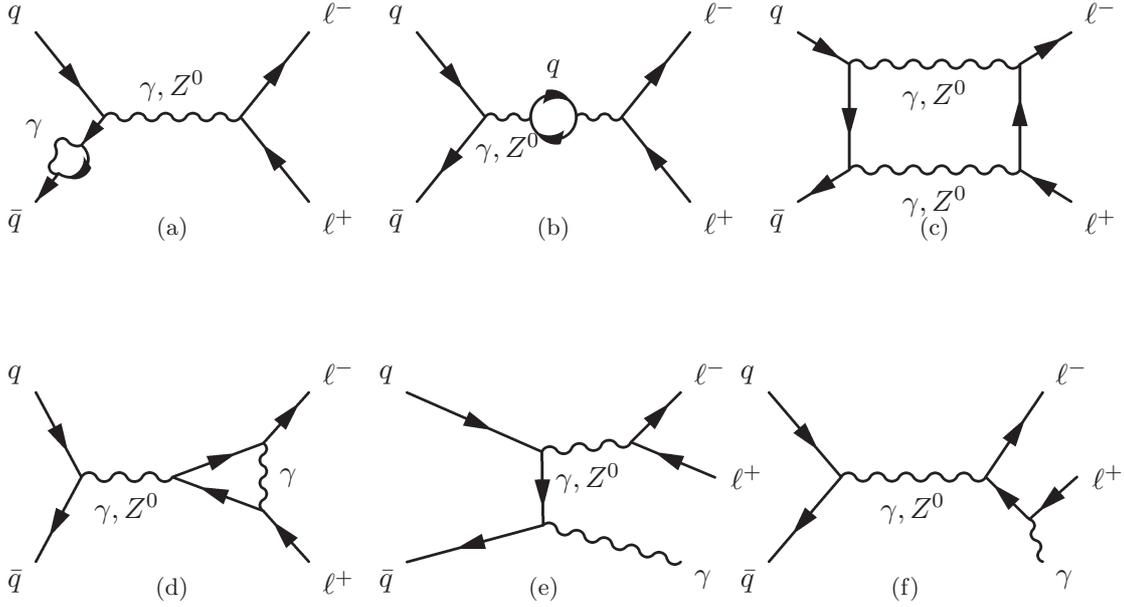
\begin{figure}

\begin{fmffile}{ewrc_dy_nc}
\newenvironment{dync}
{
  \begin{fmfgraph*}(50,25)
    \fmfleft{i2,i1}
    \fmfright{o1,o2}
    \fmflabel{$q$}{i1}
    \fmflabel{$\bar{q}$}{i2}
    \fmflabel{$\ell^+$}{o1}
    \fmflabel{$\ell^-$}{o2}
}
{
  \end{fmfgraph*}
}
\subfloat[]{\label{fig:fdiag_ewrc_dync1}
\begin{dync}
  \fmf{fermion}{i1,v1}
  \fmf{fermion,tension=3.}{v1,v3}
  \fmf{fermion,left,tension=2}{v3,v4}
  \fmf{photon,left=1,label=$\gamma$}{v4,v3}
  \fmf{fermion,tension=3.}{v4,i2}
  \fmf{fermion}{o1,v2,o2}
  \fmf{photon,label=$\gamma,,Z^0$}{v1,v2}
\end{dync}
}
\subfloat[]{\label{fig:fdiag_ewrc_dync2}
\begin{dync}
  \fmf{fermion}{i1,v1,i2}
  \fmf{boson,tension=3.,label=$\gamma,,Z^0$}{v1,v3}
  \fmf{fermion,left,tension=1.5,label=$q$}{v3,v4}
  \fmf{fermion,left,tension=1.5}{v4,v3}
  \fmf{boson,tension=3.}{v4,v2}
  \fmf{fermion}{o1,v2,o2}
\end{dync}
}
\subfloat[]{\label{fig:fdiag_ewrc_dync3}
\begin{dync}
  \fmf{fermion}{i1,v1}
  \fmf{fermion,tension=0.3}{v1,v2}
  \fmf{fermion}{v2,i2}
  \fmf{boson,tension=0.3,label=$\gamma,,Z^0$}{v1,v4}
  \fmf{boson,tension=0.3,label=$\gamma,,Z^0$}{v2,v3}
  \fmf{fermion}{o1,v3}
  \fmf{fermion,tension=0.3}{v3,v4}
  \fmf{fermion}{v4,o2}
\end{dync}
}\\ [3\baselineskip]
\subfloat[]{\label{fig:fdiag_ewrc_dync4}
\begin{dync}
  \fmf{fermion}{i1,v1,i2}
  \fmf{boson,label=$\gamma,,Z^0$}{v1,v2}
  \fmf{fermion,tension=0.5}{v3,v2,v4}
  \fmf{boson,tension=0.5,label=$\gamma$}{v3,v4}
  \fmf{fermion}{o1,v3}
  \fmf{fermion}{v4,o2}
\end{dync}
}
\subfloat[]{\label{fig:fdiag_ewrc_dync5}
\begin{fmfgraph*}(50,25)
  \fmfleft{i2,i1}
  \fmfright{o1,o2,o3}
  \fmflabel{$q$}{i1}
  \fmflabel{$\bar{q}$}{i2}
  \fmflabel{$\gamma$}{o1}
  \fmflabel{$\ell^+$}{o2}
  \fmflabel{$\ell^-$}{o3}
  \fmf{fermion}{i1,v1,v2,i2}
  \fmf{boson,tension=1.5,label=$\gamma,,Z^0$}{v1,v3}
  \fmf{fermion}{o2,v3,o3}
  \fmf{photon}{o1,v2}
\end{fmfgraph*}
}
\subfloat[]{\label{fig:fdiag_ewrc_dync6}
\begin{fmfgraph*}(50,25)
  \fmfleft{i2,i1}
  \fmfright{o1,o2,o3}
  \fmflabel{$q$}{i1}
  \fmflabel{$\bar{q}$}{i2}
  \fmflabel{$\gamma$}{o1}
  \fmflabel{$\ell^+$}{o2}
  \fmflabel{$\ell^-$}{o3}
  \fmf{fermion}{i1,v1,i2}
  \fmf{boson,label=$\gamma,,Z^0$}{v1,v2}
  \fmf{fermion}{o2,v3}
  \fmf{fermion,tension=2.}{v3,v2}
  \fmf{fermion}{v2,o3}
  \fmf{photon,tension=3.}{o1,v3}
\end{fmfgraph*}
}

\end{fmffile}
\caption{Examples of Feynman diagramms corresponding to electroweak
corrections for neutral current Drell-Yan process.}
\label{fig:fdiag_ewrc_dync}
\end{figure}

\begin{figure}

\begin{fmffile}{ewrc_dy_cc}
\newenvironment{dycc}
{
  \begin{fmfgraph*}(50,25)
    \fmfleft{i2,i1}
    \fmfright{o1,o2}
    \fmflabel{$u$}{i1}
    \fmflabel{$\bar{d}$}{i2}
    \fmflabel{$\ell^+$}{o1}
    \fmflabel{$\nu_{\ell}$}{o2}
}
{
  \end{fmfgraph*}
}
\subfloat[]{\label{fig:fdiag_ewrc_dycc1}
\begin{dycc}
  \fmf{fermion}{i1,v1}
  \fmf{fermion,tension=3.}{v1,v3}
  \fmf{fermion,left,tension=2}{v3,v4}
  \fmf{photon,left=1,label=$\gamma$}{v4,v3}
  \fmf{fermion,tension=3.}{v4,i2}
  \fmf{fermion}{o1,v2,o2}
  \fmf{photon,label=$W^+$}{v1,v2}
\end{dycc}
}
\subfloat[]{\label{fig:fdiag_ewrc_dycc2}
\begin{dycc}
  \fmf{fermion}{i1,v1,i2}
  \fmf{boson,tension=3.,label=$W^+$}{v1,v3}
  \fmf{fermion,left,tension=1.5,label=$t$}{v3,v4}
  \fmf{fermion,left,tension=1.5,label=$b$}{v4,v3}
  \fmf{boson,tension=3.}{v4,v2}
  \fmf{fermion}{o1,v2,o2}
\end{dycc}
}
\subfloat[]{\label{fig:fdiag_ewrc_dycc4}
\begin{dycc}
  \fmf{fermion}{i1,v1}
  \fmf{fermion,tension=0.3}{v1,v2}
  \fmf{fermion}{v2,i2}
  \fmf{boson,tension=0.3,label=$W^+$}{v1,v4}
  \fmf{boson,tension=0.3,label=$\gamma,,Z^0$}{v2,v3}
  \fmf{fermion}{o1,v3}
  \fmf{fermion,tension=0.3}{v3,v4}
  \fmf{fermion}{v4,o2}
\end{dycc}
}
\\ [3\baselineskip]
\subfloat[]{\label{fig:fdiag_ewrc_dycc6}
\begin{dycc}
  \fmf{fermion}{i1,v1,i2}
  \fmf{boson,label=$W^+$}{v1,v2}
  \fmf{boson,tension=0.5,label=$W^+$,l.side=left}{v3,v2}
  \fmf{boson,tension=0.5,label=$Z^0$,l.side=left}{v2,v4}
  \fmf{fermion,tension=0.5,label=$\nu_{\ell}$}{v3,v4}
  \fmf{fermion}{o1,v3}
  \fmf{fermion}{v4,o2}
\end{dycc}
}
\subfloat[]{\label{fig:fdiag_ewrc_dycc7}
\begin{fmfgraph*}(50,25)
  \fmfleft{i2,i1}
  \fmfright{o1,o2,o3}
  \fmflabel{$u$}{i1}
  \fmflabel{$\bar{d}$}{i2}
  \fmflabel{$\gamma$}{o1}
  \fmflabel{$\ell^+$}{o2}
  \fmflabel{$\nu_{\ell}$}{o3}
  \fmf{fermion}{i1,v1,v2,i2}
  \fmf{boson,tension=1.5,label=$W^+$}{v1,v3}
  \fmf{fermion}{o2,v3,o3}
  \fmf{photon}{o1,v2}
\end{fmfgraph*}
}
\subfloat[]{\label{fig:fdiag_ewrc_dycc8}
\begin{fmfgraph*}(50,25)
  \fmfleft{i2,i1}
  \fmfright{o1,o2,o3}
  \fmflabel{$u$}{i1}
  \fmflabel{$\bar{d}$}{i2}
  \fmflabel{$\gamma$}{o1}
  \fmflabel{$\ell^+$}{o2}
  \fmflabel{$\nu_{\ell}$}{o3}
  \fmf{fermion}{i1,v1,i2}
  \fmf{boson,label=$W^+$}{v1,v2}
  \fmf{fermion}{o2,v3}
  \fmf{fermion,tension=2.}{v3,v2}
  \fmf{fermion}{v2,o3}
  \fmf{photon,tension=3.}{o1,v3}
\end{fmfgraph*}
}

\end{fmffile}
\caption{Examples of Feynman diagramms corresponding to electroweak 
corrections for charged current Drell-Yan process.}
\label{fig:fdiag_ewrc_dycc}
\end{figure}
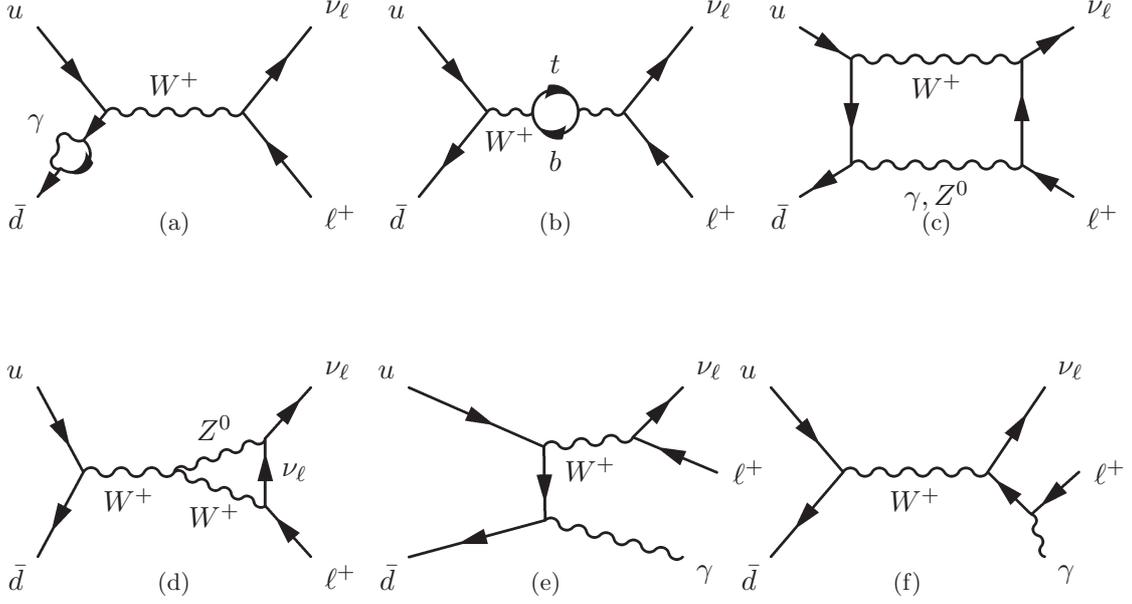

The tree level diagrams for the DY process are shown in
figure~\ref{fig:fdiag_tree_dy} for neutral and charged currents. Examples of the
diagrams corresponding to the electroweak NLO component for neutral and
charged currents are shown in figures~\ref{fig:fdiag_ewrc_dync}
and~\ref{fig:fdiag_ewrc_dycc} respectively.

For real photon emission we separate contributions from initial and
final state radiation and their interference in a gauge invariant way.
In case of photon emission off the virtual W we introduce the splitting of
the W-boson propagators by the following formula:
\begin{align}
\begin{split}
\frac{1}{\hat{s}-(M_W-i\Gamma_W)^2} &\frac{1}{\hat{s}'-(M_W-i\Gamma_W)^2}\\
=\quad &\frac{1}{(\hat{s}-\hat{s}')}
\biggl(\frac{1}{\hat{s}'-(M_W-i\Gamma_W)^2}
- \frac{1}{\hat{s}-(M_W-i\Gamma_W)^2}\biggr).
\end{split}
\end{align}

The so-called {\it on-shell} singularities which appear in form of
logarithms $\log(\hat{s}-M_W^2+i\epsilon)$ can be regularized by the $W$-width
\cite{Wackeroth:1996hz}:
\begin{align}
\log(\hat{s}'-M_W^2+i\epsilon) \to \log(\hat{s}'-M_W^2+iM_W\Gamma_W).
\end{align}

\unitlength = 0.95mm

\unitlength = 1pt

\unitlength = 0.9mm

\unitlength = 1pt

Electroweak NLO radiative corrections contain terms proportional to logarithms of
the quark masses, $\log(\hat{s}/m_{u,d}^2)$.  They come from the initial state
radiation contributions including hard, soft and virtual photon emission. Such
initial state mass singularities are well known, for instance, in the process
of $e^+e^-$ annihilation. However, in the case of hadron collisions these logs have
been already {\em effectively} taken into account in the parton density
functions (PDF).  In fact, in the procedure of PDF extraction from 
experimental data, the QED radiative corrections to the quark line were not
systematically subtracted. Therefore current PDFs effectively include not
only the QCD evolution but also the QED one.  Moreover, it is known that the
leading log behaviours of the QED and QCD DGLAP evolution of the quark density
functions are similar (proportional to each other). Consequently one gets an evolution of
the PDF with an effective coupling constant
\begin{align}
\alpha^{\mathrm{eff}}_{s} \approx \alpha_{s} + \frac{Q_i^2}{C_F}\alpha,
\end{align}
where $\alpha_s$ is the strong coupling constant, $\alpha$ is the fine
structure constant, $Q_i$ is the quark charge, and $C_F$ is the QCD colour
factor.

We will use here the \MSbar subtraction scheme, the DIS scheme may be used as
well. A solution described in \cite{Diener:2005me} allows to avoid 
the double counting of the initial quark mass
singularities contained in our result for the corrections to the free quark
cross section and the ones contained in the corresponding PDF.  The latter
should also be taken in the same scheme with the same factorization scale.

The \MSbar subtraction to the fixed (leading) order in $\alpha$ is given by:
\begin{align}
\label{msbarq}
\begin{split}
\bar{q}(x,M^2)& = q(x,M^2) -
\int_x^1 \frac{\mathrm{d} z}{z} \, q\biggl(\frac{x}{z},M^2\biggr) \,
\frac{\alpha}{2\pi} \, Q_q^2
\biggl[ \frac{1+z^2}{1-z}
\biggl\{\ln\biggl(\frac{M^2}{m_q^2}\biggr)-2\ln(1-z)-1\biggr\} \biggr]_+
\\ 
& \equiv q(x,M^2) - \Delta q,
\end{split}
\end{align}
where \(q(x,M^2)\) is the parton density function in the \MSbar scheme computed
using the QED DGLAP evolution.

The differential hadronic cross section for DY processes~[\ref{eqn:dy_process}]
is given by
\begin{align}
  \mathrm{d}\sigma^{pp \to \ell\ell'X} = \sum_{q_{1}q_{2}} \int\limits_0^1 \int\limits_0^1 
  \mathrm{d}x_1 \, \mathrm{d}x_2 \, \bar{q}_1(x_1,M^2) \, \bar{q}_2(x_2,M^2)\,\mathrm{d}\hat{\sigma}^{q_1 q_2\to \ell\ell'},
\end{align}
where \(\bar{q}_1(x_1,M^2), \bar{q}_2(x_2,M^2) \) are the parton density
functions of the incoming quarks modified by the subtraction of the quark mass
singularities and \( \sigma^{q_1 q_2\to \ell\ell'} \) is the partonic cross
section of corresponding hard process.  The sum is performed over all
possible quark combinations for a given type of process ($q_1q_2 = u\bar{d},
u\bar{s}, c\bar{d}, c\bar{s}$ for CC and $q_1q_2 = u\bar{u}, d\bar{d},
s\bar{s}, c\bar{c}, b\bar{b}$ for NC). In our calculations we used fixed
factorization scales $M^2 = M_W^2$ for CC and $M^2 = M_Z^2$ for NC.

The effect of applying different EW schemes in the SANC system is discussed
in~\cite{Arbuzov:2007db}. In the current study we are using the
\(G_{\mu}\)-scheme~\cite{Degrassi:1996ps} since it minimizes EW radiative
corrections to the inclusive DY cross section.  In this scheme the weak coupling
\(g\) is related to the Fermi constant and the W boson mass by equation
\begin{align}
g^2 = 4\sqrt{2} G_{\mu} m_{W}^2(1-\Delta r),
\label{eqn:coupling}
\end{align}
where \(\Delta r\) represents all radiative corrections to the muon decay
amplitude~\cite{Sirlin:1980nh}. Since the vertex term between charged particles and
photons is proportional to \(g \sin{\theta_W}\), one can introduce an effective
electromagnetic coupling constant
\begin{align}
\alpha_{G_{\mu}}^{tree} = \frac{\sqrt{2} G_{\mu} \sin^2\theta_W m_W^2}{\pi},
\end{align}
which is evaluated from~(\ref{eqn:coupling}) in a tree-level approximation by
setting \(\Delta r = 0 \). 

The total NLO electroweak corrections to the total charged current DY cross
section for 14~TeV \(pp\) collisions are estimated to be about \(-2.7\%\) for
the \(G_{\mu}\)-scheme and can reach up to \(10\%\) for the differential cross section
in certain kinematical regions~\cite{Arbuzov:2005dd, Arbuzov:2007db}.

The EW NLO calculations for the DY processes were performed using
semi-analytic calculations with the aid of the FORM symbolic manipulation
system~\cite{Vermaseren:2000nd} and employ LoopTools~\cite{Hahn:1998yk} and
SancLib~\cite{Andonov:2004hi} libraries for evaluation of scalar and tensor
one-loop integrals. The analytical expressions for different components of the
differential EW NLO cross-section for DY processes are realized within
standard {\tt SANC} Fortran modules which are used in our Monte Carlo
event generators of unweighted events.

\subsection{Photon induced contributions}

At the ${\mathcal O}(\alpha)$ level one can see that there is a non-zero
probability to find a quasi-real photon inside one of the colliding protons.
This brings up an additional QED contribution to the EW corrections, so called
inverse bremsstrahlung. The complete set of ${\mathcal O}(\alpha)$
photon-induced contributions for both NC and CC Drell--Yan processes was
evaluated in~\cite{Arbuzov:2007kp}. The charged current results for this
component were given by S.~Dittmaier and M.~Kr\"{a}mer in the proceedings to
the Les Houches workshop~\cite{Buttar:2006zd}. 
The results for the neutral current
were presented in~\cite{Dittmaier:2009cr}, using an approach 
which implies effective resummation
of the top quark one- and two-loop corrections in the LO cross section via \(\stw\) renormalization:
\begin{align}
\stws \to \stwsb \equiv \stws+\ctws\Delta\rho, \qquad
\ctws \to \ctwsb \equiv \ctws(1-\Delta\rho),
\end{align}
where
\begin{align}
\Delta\rho = \Delta\rho^{(1)}[1+\rho^{(2)}(M_H^2/\mtp^2)\Delta\rho^{(1)}/3][1-\frac{2\alpha_s}{9\pi}(\pi^2+3)]
\end{align}
with \(\Delta\rho^{(1)} \propto G_{\mu}\mtp^2\)~\cite{Denner:1991kt} and with the function \(\rho^{(2)}\) given
in~\cite{Fleischer:1993ub}. The coupling constant \(\alpha_{G_{\mu}}\) is replaced by \(\alpha_{G_{\mu}}\stwsb/\stws\)
in this approach.
Table~\ref{tab_nc_comparison} presents comparison of SANC results with~\cite{Dittmaier:2009cr} 
with corresponding input parameters for the photon-induced 
contribution without these key 
differences in calculation schemes taken into account. In SANC the
corresponding  
effects are considered as a part of the first- (and higher) order radiative
corrections.

This comparison shows that although the size of the photon-induced contribution
can differ by 10\% or more between the two approaches, this corresponds to at
most a per mille level difference in the total lepton pair cross section. It is
therefore smaller than the aimed for accuracy and we do not need to consider
this difference in further detail.

\begin{figure}[ht]
\[
\scalebox{0.7}{
\begin{picture}(440,290)(0,0)
\ArrowLine(20,20)(100,40)
\Vertex(100,40){2}
\ArrowLine(100,40)(180,20)
\Photon(100,40)(100,80){3}{5}
\Vertex(100,80){2}
\ArrowLine(100,80)(100,120)
\ArrowLine(180,80)(100,80)
\Photon(20,140)(100,120){3}{6}
\Vertex(100,120){2}
\ArrowLine(100,120)(180,140)
\Text(30,150)[]{\Large $ \gamma $}
\Text(30,10)[]{\Large $ q $}
\Text(170,12)[]{\Large $ q $}
\Text(170,70)[]{\Large $ \ell^+ $}
\Text(170,150)[]{\Large $ \ell^- $}
\Text(115,100)[]{\Large $ \ell^- $}
\Text(82,60)[]{\Large $ \gamma,Z $}

\ArrowLine(260,20)(340,40)
\Vertex(340,40){2}
\ArrowLine(340,40)(420,20)
\Photon(340,40)(340,80){3}{5}
\Vertex(340,80){2}
\ArrowLine(340,120)(340,80)
\ArrowLine(340,80)(420,80)
\Photon(260,140)(340,120){3}{6}
\Vertex(340,120){2}
\ArrowLine(420,140)(340,120)
\Text(270,150)[]{\Large $ \gamma $}
\Text(270,10)[]{\Large $ q $}
\Text(410,12)[]{\Large $ q $}
\Text(410,70)[]{\Large $ \ell^- $}
\Text(410,150)[]{\Large $ \ell^+ $}
\Text(355,100)[]{\Large $ \ell^+ $}
\Text(322,60)[]{\Large $ \gamma,Z $}

\Photon(20,280)(70,240){3}{6}
\ArrowLine(20,200)(70,240)
\Vertex(70,240){2}
\ArrowLine(70,240)(120,240)
\Vertex(120,240){2}
\ArrowLine(120,240)(170,280)
\Photon(120,240)(150,216){3}{5}
\Vertex(150,216){2}
\ArrowLine(200,246)(150,216)
\ArrowLine(150,216)(200,186)
\Text(30,290)[]{\Large $ \gamma $}
\Text(30,195)[]{\Large $ q $}
\Text(162,290)[]{\Large $ q $}
\Text(190,180)[]{\Large $ \ell^- $}
\Text(190,255)[]{\Large $ \ell^+ $}
\Text(95,255)[]{\Large $ q $}
\Text(120,218)[]{\Large $ \gamma,Z $}

\ArrowLine(260,200)(340,220)
\Vertex(340,220){2}
\Photon(340,220)(380,210){3}{5}
\ArrowLine(340,220)(340,260)
\Photon(260,280)(340,260){3}{6}
\Vertex(340,260){2}
\ArrowLine(340,260)(420,280)
\Vertex(380,210){2}
\ArrowLine(440,240)(380,210)
\ArrowLine(380,210)(440,180)
\Text(270,290)[]{\Large $ \gamma $}
\Text(270,190)[]{\Large $ q $}
\Text(410,290)[]{\Large $ q $}
\Text(430,175)[]{\Large $ \ell^- $}
\Text(430,250)[]{\Large $ \ell^+ $}
\Text(352,240)[]{\Large $ q $}
\Text(355,200)[]{\Large $ \gamma,Z $}
\end{picture}}
\]
\caption {Feynman diagrams for inverse bremsstrahlung in the neutral current Drell--Yan
sub-process.}
\label{fig:fdiag_inbr_nc}
\end{figure}
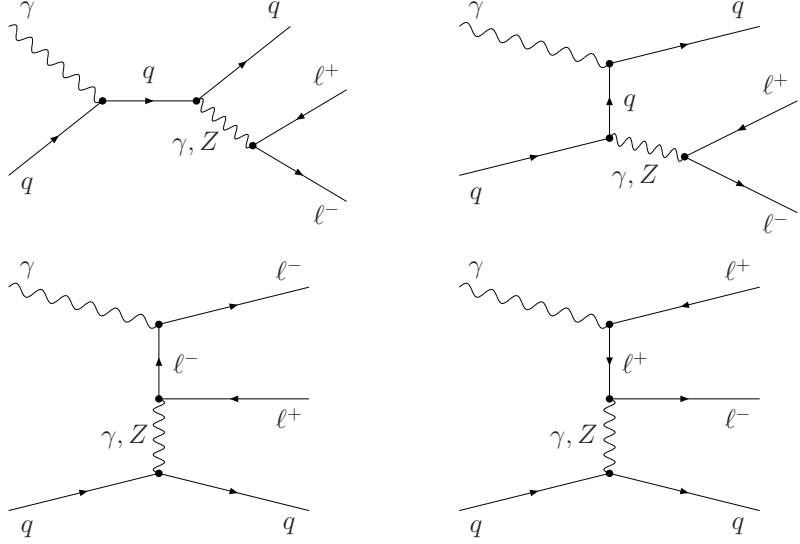

The fixed-order diagrams corresponding to the processes
\begin{align}
\begin{split}
\gamma &+ q \to q' + \ell^+ + \nu_\ell,\\
\gamma &+ q \to q + \ell^- + \ell^+
\end{split}
\end{align}
are shown in figures~\ref{fig:fdiag_inbr_cc} and~\ref{fig:fdiag_inbr_nc}. 
The inverse bremsstrahlung component for the hadronic cross section can be written in a standard way:
\begin{align}
\begin{split}
\dd\sigma_{\mathrm{inv.brem.}}^{pp\to\ell\bar{\ell}'X} &= \sum\limits_{q}\int\limits_{0}^{1}\int\limits_{0}^{1}\dd x_1\,\dd x_2\,
f_q(x_1,M^2) \, f_\gamma(x_2,M^2)\,\dd\hat{\sigma}^{q\gamma\to q'\ell\bar{\ell}'},
\end{split}
\end{align}
where \(f_q(x_1,M^2)\) and \( f_\gamma(x_2,M^2) \) are the parton density
functions for quark and photon respectively. The quark mass singularity
subtraction is performed for this contribution in analogous way to the
processes with two quarks in the initial state.

\begin{figure}[ht]
\[
\scalebox{0.7}{
\begin{picture}(440,290)(0,0)
\ArrowLine(20,20)(100,40)
\Vertex(100,40){2}
\ArrowLine(100,40)(180,20)
\Photon(100,40)(100,80){3}{5}
\Photon(20,100)(100,80){3}{6}
\Vertex(100,80){2}
\Photon(100,80)(140,90){3}{6}
\Vertex(140,90){2}
\ArrowLine(200,120)(140,90)
\ArrowLine(140,90)(200,60)
\Text(30,110)[]{\Large $ \gamma $}
\Text(30,10)[]{\Large $ q $}
\Text(170,12)[]{\Large $ q' $}
\Text(190,55)[]{\Large $ \nu_\ell $}
\Text(190,130)[]{\Large $ \ell^+ $}
\Text(122,100)[]{\Large $ W^+ $}
\Text(85,60)[]{\Large $ W^+ $}

\ArrowLine(260,20)(340,40)
\Vertex(340,40){2}
\ArrowLine(340,40)(420,20)
\Photon(340,40)(340,80){3}{5}
\Vertex(340,80){2}
\ArrowLine(340,120)(340,80)
\ArrowLine(340,80)(420,80)
\Photon(260,140)(340,120){3}{6}
\Vertex(340,120){2}
\ArrowLine(420,140)(340,120)
\Text(270,150)[]{\Large $ \gamma $}
\Text(270,10)[]{\Large $ q $}
\Text(410,12)[]{\Large $ q'$}
\Text(410,70)[]{\Large $ \nu_\ell $}
\Text(410,150)[]{\Large $ \ell^+ $}
\Text(355,100)[]{\Large $ \ell^+ $}
\Text(355,60)[]{\Large $ W^+ $}

\Photon(20,280)(70,240){3}{6}
\ArrowLine(20,200)(70,240)
\Vertex(70,240){2}
\ArrowLine(70,240)(120,240)
\Vertex(120,240){2}
\ArrowLine(120,240)(170,280)
\Photon(120,240)(150,216){3}{5}
\Vertex(150,216){2}
\ArrowLine(200,246)(150,216)
\ArrowLine(150,216)(200,186)
\Text(30,290)[]{\Large $ \gamma $}
\Text(30,195)[]{\Large $ q $}
\Text(162,290)[]{\Large $ q' $}
\Text(190,180)[]{\Large $ \nu_\ell $}
\Text(190,255)[]{\Large $ \ell^+ $}
\Text(95,255)[]{\Large $ q $}
\Text(122,218)[]{\Large $ W^+ $}

\ArrowLine(260,200)(340,220)
\Vertex(340,220){2}
\Photon(340,220)(380,210){3}{5}
\ArrowLine(340,220)(340,260)
\Photon(260,280)(340,260){3}{6}
\Vertex(340,260){2}
\ArrowLine(340,260)(420,280)
\Vertex(380,210){2}
\ArrowLine(440,240)(380,210)
\ArrowLine(380,210)(440,180)
\Text(270,290)[]{\Large $ \gamma $}
\Text(270,190)[]{\Large $ q $}
\Text(410,290)[]{\Large $ q' $}
\Text(430,175)[]{\Large $ \nu_\ell $}
\Text(430,250)[]{\Large $ \ell^+ $}
\Text(354,242)[]{\Large $ q' $}
\Text(355,200)[]{\Large $ W^+ $}
\end{picture}}
\]
\caption {Feynman diagrams for inverse bremsstrahlung in the charged current Drell--Yan
sub-process.}
\label{fig:fdiag_inbr_cc}
\end{figure}
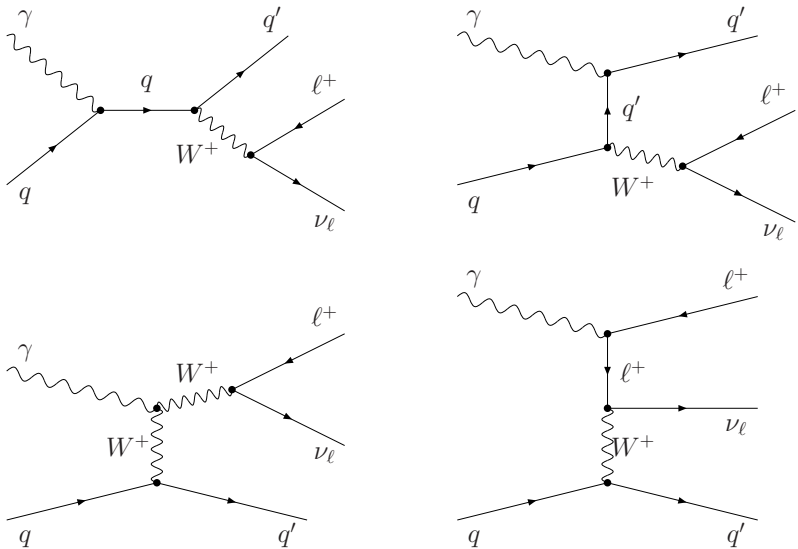

\begin{figure}[h]
\subfloat[]{\label{fig:transv_mass_munu}
\includegraphics*[width=0.5\textwidth]{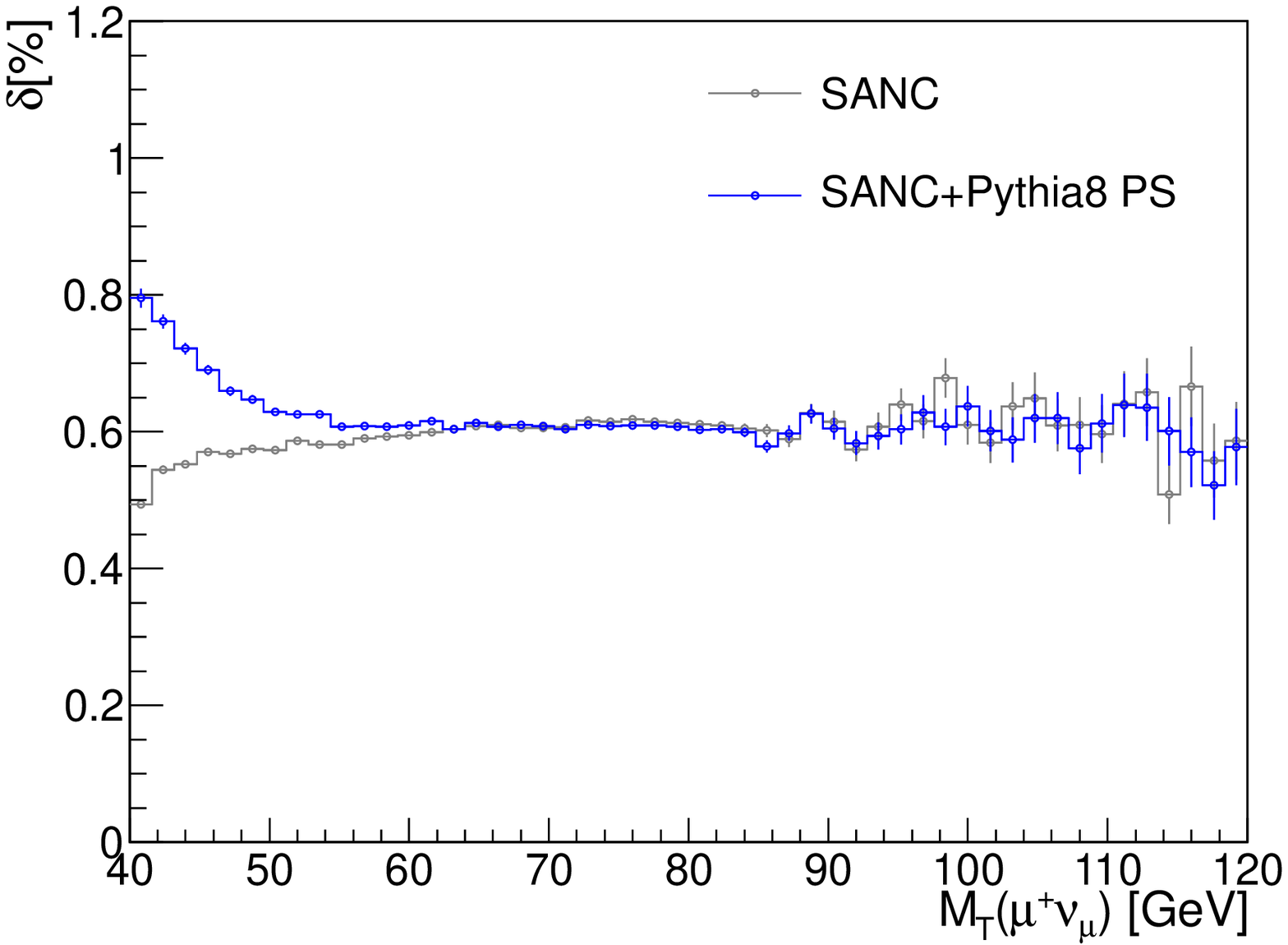}\\
}
\subfloat[]{\label{fig:transv_mom_mupl}
\includegraphics*[width=0.5\textwidth]{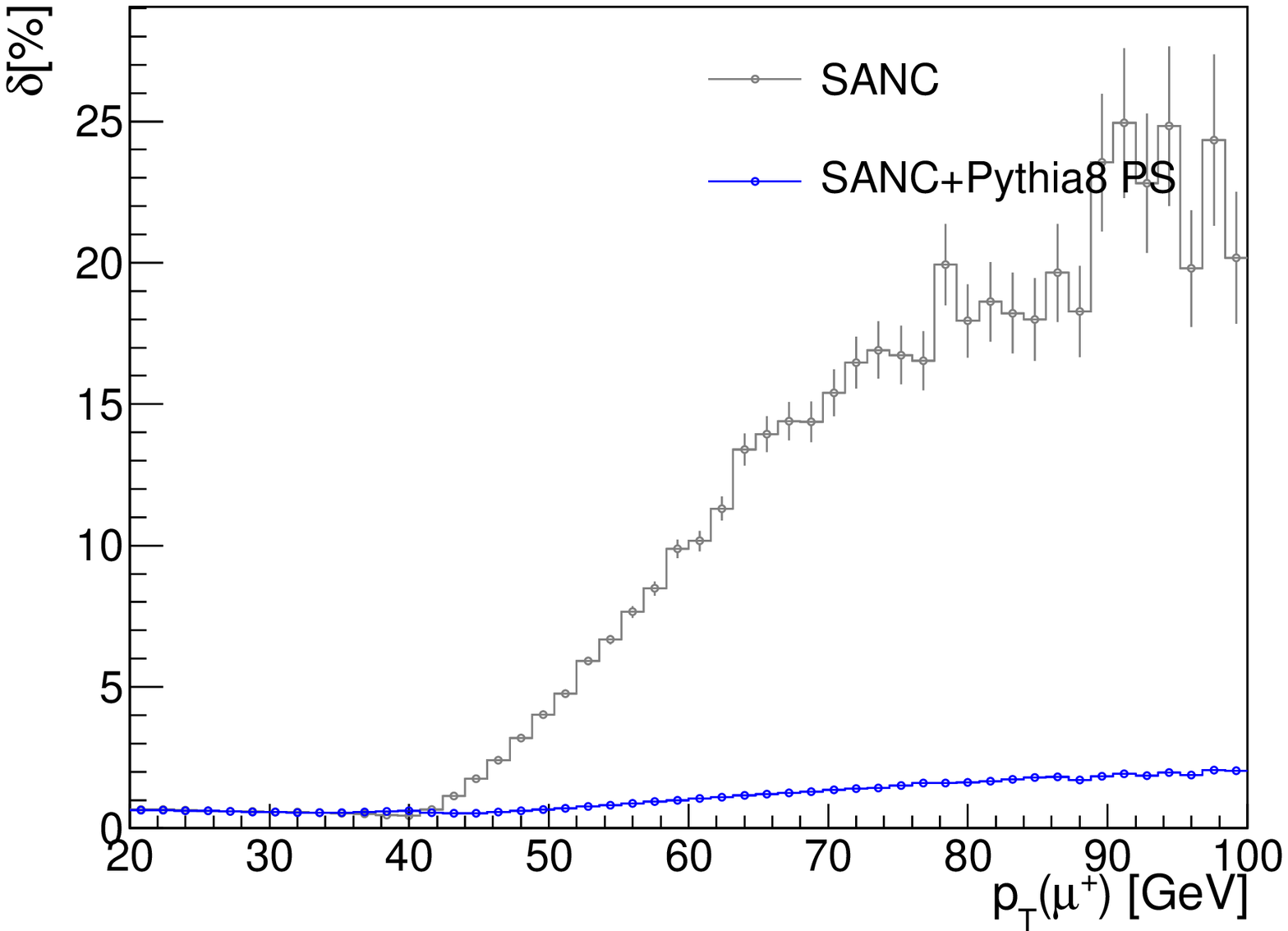}\\
}
\caption {Distributions of the inverse bremsstrahlung contribution correction \(
\delta_{\gamma q} \) for muon-neutrino pair transverse mass~(a) and muon
transverse momentum~(b) in the charged current Drell--Yan process.}
\label{fig:delta_distr} 
\end{figure}

\begin{figure}[h]
\subfloat[]{\label{fig:inv_mass_mumu}
\includegraphics*[width=0.5\textwidth]{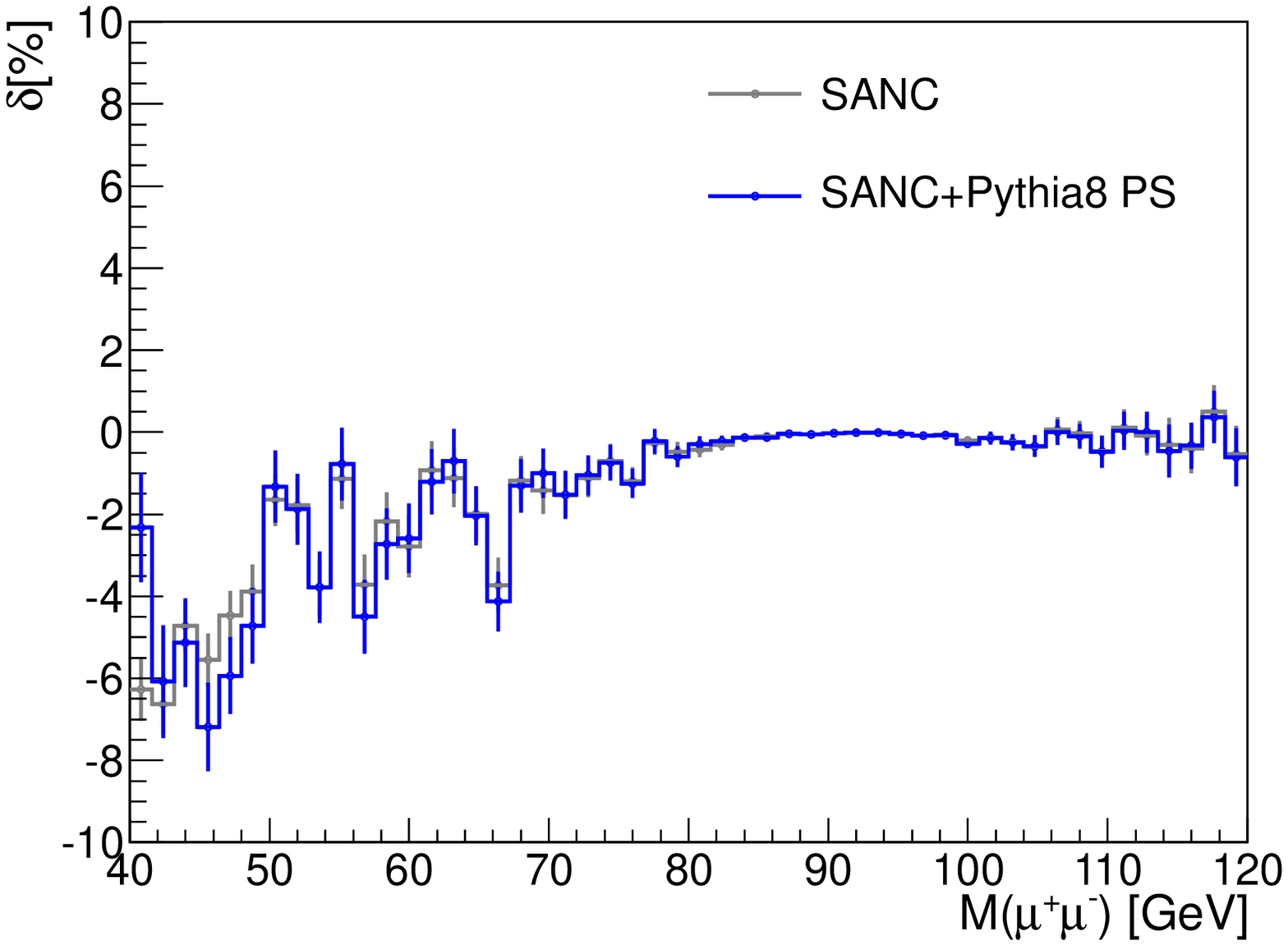}\\
}
\subfloat[]{\label{fig:transv_mom_mupl_nc}
\includegraphics*[width=0.5\textwidth]{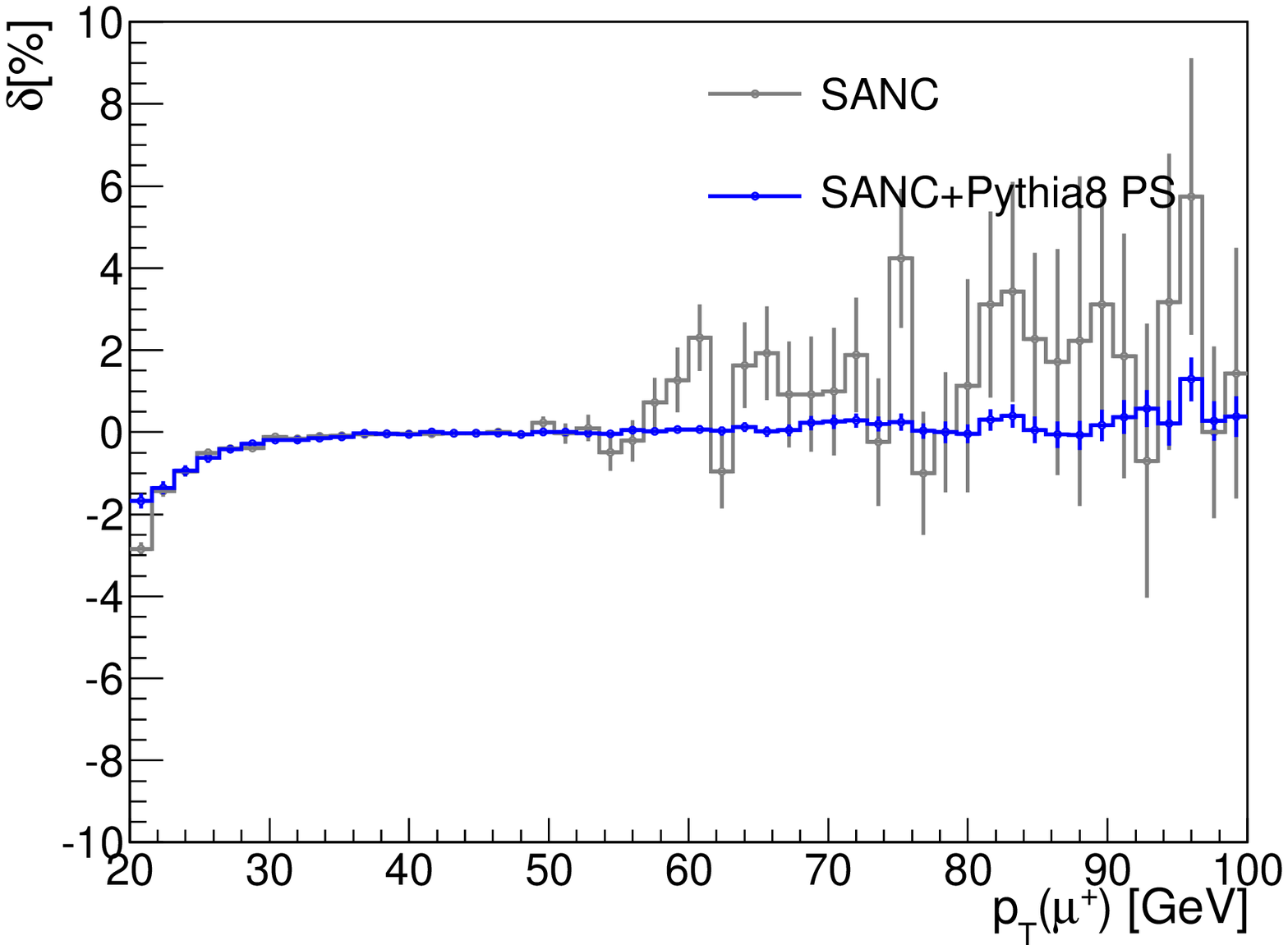}\\
}
\caption {Distributions of the inverse bremsstrahlung contribution correction \(
\delta_{\gamma q} \) for $\mu^+ \mu^-$ pair invariant mass~(a) and muon
transverse momentum~(b) in the neutral current Drell--Yan process.}
\label{fig:delta_distr_nc} 
\end{figure}

The photon induced channels are explicitly included in the SANC event
generators. The corresponding correction value defined as \(\delta_{\gamma q} =
\sigma_{\gamma q}/\sigma_0 \), where \(\sigma_0\) is the tree level process
cross section, is below the percent level for the total cross section, but reaches
several percents in certain kinematic regions. The corrections for
muon-neutrino pair transverse mass and \(\mu^+\) transverse momentum
distributions in the charged current process \( pp\to \mu^+ \nu \) for \(
\delta_{\gamma q} \) are shown in figure~\ref{fig:delta_distr}.
The corrections for \(\mu^+\mu^-\) invariant mass and \(\mu^+\) transverse momentum
distributions in the neutral current process \( pp\to \mu^+ \mu^- \) for \(
\delta_{\gamma q} \) are shown in figure~\ref{fig:delta_distr_nc}. The large
corrections for \(\mu^+\) transverse momentum in the charged current process
in the region of \(p_T(\mu^+) > M_W/2\) is due to the recoil of a virtual $W$.

\begin{table}[t]
\begin{center}
\begin{tabular}{|l|l|l|l|l|l|l|}
  \hline
  $ \mathrm{M_{min} (\mu^+\mu^-), GeV} $ & 50 & 100 & 200 & 500 & 1000 & 2000\\
  \hline
  $ \mathrm{\sigma_{Born}^{\tt SANC}, pb} $ & 739.21 & 32.880 & 1.4874 & 0.081078 & 0.0068107 & 0.00030423\\
  $ \mathrm{\sigma_{Born}^{\tt DH}, pb} $ & 738.73 & 32.724 & 1.4848 & 0.080942 & 0.0067995 & 0.00030374\\
  \hline
  $ \mathrm{\delta_{\gamma q}^{\tt SANC}, \%} $ & -0.11(1) & -0.21(1) & 0.46(1) & 1.68(1) & 2.05(1) & 2.26(1)\\
  $ \mathrm{\delta_{\gamma q}^{\tt DH}, \%} $ & -0.11(1) & -0.21(1) & 0.38(1) & 1.53(1) & 1.91(1) & 2.34(1)\\
  \hline
\end{tabular}
\end{center}
\caption {The $\sigma_{Born}$ cross-sections and corresponding corrections $\delta_{\gamma q}$,
obtained in SANC and S. Dittmaier and M. Huber (DH) for the neutral current process}
\label{tab_nc_comparison}
\end{table}

The inverse bremsstrahlung contributions can be of resonant and non-resonant
type. The latter have the incoming photon coupling to leptons and require
a special colour flow interpretation in the code used to apply QCD parton 
showers to the hard process. As a workaround one can write an event entry
for such contributions as a \(2\to3\) process without internal structure.
The resonant component can be treated in a standard way indicating
a Z boson as a virtual propagator.

\section{Parton Showers \label{psmc}}

In contrast to the fixed-order calculations described above, parton showers
rely on an iterative (Markov-chain) branching procedure to reach arbitrary
orders in the perturbative expansion. By keeping the total normalization
unchanged, the shower explicitly conserves unitarity at each order,  generating
equal-magnitude but opposite-sign real and virtual corrections. Each branching
step is based on universal splitting functions that capture the leading
singularities of the full higher-order matrix elements exactly. Subleading
terms can usually only be taken into account approximately, and hence different
shower models (and ``tunings'') can give different answers outside the strict
soft/collinear limits. 

Still, in practice, parton showers are reasonably accurate  even for finite
emission energies and angles, as long as the characteristic scale of each
emission is hierarchically smaller than that of the preceding process (strong
ordering). As such, they are complementary to the fixed-order truncations
discussed above, which are accurate only in the absence of large hierarchies. 

Several different shower formulations have been developed. In Herwig++ and
Pythia8, which we shall be concerned with here, the shower approximation is
cast in terms of evolution equations using DGLAP splitting kernels, which
nominally capture only the leading-logarithmic (LL)  behaviour of higher
perturbative orders.  To further improve the accuracy, parton showers
incorporate a number of improvements relative to the naive
leading-log picture; 1)~they use renormalization-group improved couplings by
shifting the argument of $\alpha_s$ for shower emissions to
$\alpha_s(p_\perp)$, thereby absorbing the $\beta$-function-dependent terms in
the one-loop splitting functions into the effective tree-level ones, 2)~they
approximately  incorporate the higher-order interference effect known as
coherence by imposing angular ordering either at the level of the evolution
variable (Herwig++) or in the construction of the shower phase space (Pythia8),
3)~they enforce exact momentum conservation to all  orders, albeit in different
ways between the two (different ``recoil strategies''),  and 4)~both programs
include at least some further corrections due to polarization effects. The
resulting approximation is thus significantly better than ``pure LL'', although
it cannot be formally claimed to reach the NLL level. 
 
Prior to the writing of this paper, the initial-state showers in both Herwig++
and Pythia8 included photons as emitted particles, but not as evolving ones.
Interfacing of the photon-induced subprocesses to Herwig++ and Pythia8
therefore required a certain modification of these codes. In the two following
subsections, we briefly summarize the main properties of these modifications,
for each program respectively. 

\subsection{Processes with incoming photons in Pythia8} For the Pythia8
implementation of incoming photons, we  re-use the existing backwards-evolution
framework for gluons, with the modification that there is no photon
self-coupling and replacing the $q\to g$ coupling and colour factor by
$\bar{\alpha}=e_q^2\alpha_\mathrm{em}/2\pi$ for $q \to \gamma$. For future
reference, we here summarize the steps specific to the photon
backwards-evolution.

Denoting the Pythia8 evolution variable $p_{\perp\mathrm{evol}}^2$ (see
\cite{Sjostrand:2004ef}), the evolution equation for a photon-in-a-proton is
cast as a standard Sudakov evolution, with subsequent $p_{\perp\mathrm{evol}}$
``trial'' scales generated according to an overestimate of the physical
evolution probability, obtained by solving the trial evolution equation,
\begin{equation}
R = \hat{\Delta}(p^2_{\perp\mathrm{now}}, p^2_{\perp\mathrm{next}}) =
\exp\left[ 
- \int_{p^2_{\perp\mathrm{next}}}^{p^2_{\perp\mathrm{now}}}
\frac{\mathrm{d}p_{\perp\mathrm{evol}}^2}
{p_{\perp\mathrm{evol}}^2} \ \bar{\alpha} \hat{I}_z,
\right]\label{eq:evol}
\end{equation}
where $\hat{\Delta}$ is the trial shower Sudakov, representing a lower bound on
the probability that there are no branchings between the two scales
$p_{\perp\mathrm{now}}^2$ and $p_{\perp\mathrm{next}}^2$, $R$ is a random
number distributed uniformly between zero and one, and the EM coupling used for
the trial emission is $\bar{\alpha} = \bar{\alpha}(\hat{s})$, with $\hat{s}$
being the CM energy of the two incoming partons (specifically, it is an
overestimate of $\alpha(p_\perp^2)$, which will be imposed by veto, below).
The trial $z$ integral, $\hat{I}_z$, is defined by 
\begin{equation}
\hat{I}_z = 4 \left( \frac{1}{\sqrt{z_{\mathrm{min}}}} -
\frac{1}{\sqrt{z_\mathrm{max}}}\right) \frac{\sum_i
  e_i^2f_i(x_\gamma,p_{\perp\mathrm{now}}^2)}
     {f_\gamma(x_\gamma,p_{\perp\mathrm{now}}^2)}~, \label{eq:Iz}
\end{equation}
where $x_\gamma$ is the momentum fraction carried by the incoming photon and
the $z$ limits are defined in \cite{Sjostrand:2004ef}.

Solving eq.~(\ref{eq:evol}) for $p^2_{\perp\mathrm{next}}$, we get
\begin{equation}
p_{\perp\mathrm{next}}^2 = p_{\perp\mathrm{now}}^2 \ R^{(\bar{\alpha} \hat{I}_z)^{-1}}~.
\end{equation}
Given a trial $p_{\perp\mathrm{next}}$ value obtained from this equation, we
may now generate a corresponding trial $z$ value  according to 
\begin{equation}
z = \frac{z_\mathrm{min}z_{\mathrm{max}}}{(
 \sqrt{z_\mathrm{min}} + (\sqrt{z_\mathrm{max}}-\sqrt{z_\mathrm{min}}\,)R'
)^2}~,
\end{equation}
where $R'$ is a second random number distributed uniformly between zero and
one, and the $z$ limits are the same as those used in eq.~(\ref{eq:Iz}). We
have now obtained an importance-sampled pair of trial
$(p_{\perp\mathrm{next}},z)$ values. The quark flavour $q$ is chosen with probability proportional to $e_q^2f_q(x_\gamma/z,p_{\perp\mathrm{next}}^2)$.
Since the overestimates used for the
importance sampling are not quite identical to the physical distributions we
wish to obtain, the second step of the algorithm is to employ the veto
algorithm and accept only those trials that lie inside the physical phase space
with a probability, 
\begin{equation}
P \ = \ \frac{\bar{\alpha}(p_{\perp\mathrm{next}}^2)}{\bar{\alpha}(\hat{s})} \
 \frac{f_\gamma(x_\gamma,p_{\perp\mathrm{now}}^2)}
      {f_\gamma(x_\gamma,p_{\perp\mathrm{next}}^2)} \
\frac{ f_q(x_\gamma/z,p_{\perp\mathrm{next}}^2)}
     { f_q(x_\gamma,p_{\perp\mathrm{now}}^2)} \
 \frac12 \left( 
1  + (1-z)^2\right) \sqrt{z}
 ~,\label{eq:Paccept}
\end{equation}
where the coupling factor translates the argument of $\alpha_\mathrm{em}$ in
the manner mentioned above, the first PDF factor corrects the factorization
scale used in the photon PDF to the new evolution scale, the second PDF factor
corrects both the $x$ and $Q^2$ of the would-be parent quark (or antiquark) to
their correct post-branching values, and the last factor (the $z$ factor)
corrects for the form of $P(z)$ used for the trial generation (the factor
$\sqrt{z}$, which may seem to complicate matters unnecessarily, arises since we
use a factor $1/\sqrt{z}$ in the trial generation to suppress the high-$x$ bump
on valence quark distributions, which could otherwise lead to the trial
generator not overestimating the physical splitting probability). 

If no acceptable branching is found above the global initial-state shower
cutoff (cf.~the documentation of Pythia8's spacelike showers
\cite{Sjostrand:2004ef}), the photon is considered as having been extracted
directly from the beam remnant. Also note that the maximum expressed by
eq.~(\ref{eq:Iz}) could be violated, yielding $P > 1$ in
eq.~(\ref{eq:Paccept}), if the photon PDF exhibits any thresholds or other
sharp features.  Further work would be needed to properly take into account
such structures. We note, however, that the code forces the PDF to be bounded
from below, so that a vanishing PDF should result in warnings, not crashes. 

The (yet higher order) possibility of a fermion backwards-evolving to a photon
has not yet been included in this framework. The net effect is therefore only
to allow the initial-state shower off an incoming photon to reconstruct back to
a quark or antiquark, but not the other way around. 

\subsection{Processes with incoming photons in Herwig++}

The physics implementation of processes with incoming photons is very similar
in Herwig++ to that already described for Pythia8, so we do not go into as much
detail.  However, the practical implementation is somewhat different.

When the Herwig++ parton shower is presented with a hard process with an
incoming photon, it calls a \texttt{PreShowerHandler} of the
\texttt{IncomingPhotonEvolver} class, which is specially written for this
purpose.  It generates a step of backward evolution from the photon to an
incoming quark, in exactly the same way as described above.  In particular, the
transverse momentum of the $q\to q\gamma$ vertex is required to be smaller than
the scale of the hard process, as in Pythia8.  However, this backward evolution
step is required to be generated with a probability of unity and if no backward
step is generated above the infrared cutoff, or if it is generated outside the
allowed phase space, then the evolution scale is reset to the hard scale and it
loops back to try again.  In a very small fraction of events it can happen
that, due to mismatch between the hard process (SANC) and parton shower
(e.g.~in parton mass values and hadron remnant treatment) no backward step is
possible.  In such cases an \texttt{EventException} is thrown. 

Having generated a backward step, the \texttt{IncomingPhotonEvolver} modifies
the hard process to include it.  That is, it replaces a $\gamma\to X$ event by
the corresponding $q\to qX$ event, which the rest of Herwig++'s parton shower
machinery operates on as normal.  The quark line is correctly labeled as
colour-disconnected from the rest of the hard process, so the colour coherence
inherent to Herwig++'s shower ensures that it only radiates with opening angles
smaller than the $q\to q$ scattering angle. 

The \texttt{IncomingPhotonEvolver} has one parameter that may be of interest to
users: \texttt{minpT}, the minimum transverse momentum generated for the $q\to
q\gamma$ vertex.  All of the plots shown below were generated with the default
value of 2.0~GeV. 

\section{Cross Checks and Validation \label{Scheme_sect}}

Several cross check simulations were performed in order to verify the new
implementation for the processes with incoming photons.  The simulations
included two steps: \textit{i}) hard event generation using the SANC MC
generator for charged (CC) and neutral (NC) current cases, and \textit{ii})
addition of the parton showers using Herwig++ or Pythia8 generators. At the
generation step the event selection shown in table~\ref{tab_cuts1} was applied.
\begin{table}[t]
\centerline{
\begin{tabular}{l|ll}
  \hline
  process & \multicolumn{2}{c}{ cuts} \\
  \hline
  \multirow{2}{*}{$pp\to W^{+}\to l^{+}\nu_{l}(\gamma)+X$}
      & $M_{inv}(\mu^{+} \nu_{\mu}) > 20$~GeV, 
      & $|\eta(\mu^{+})| < 4.0$, \\
      & no $p_T$ cut &\\
  \hline
  \multirow{2}{*}{$pp\to Z/\gamma\to l^+l^-(\gamma)+X$}
      & $M_{inv}(\mu^{+}\mu^{-} ) > 20$~GeV,
      & $|\eta(\mu^{\pm})| < 3.5$, \\
      & $p_T(\mu^{\pm}) > 2$~GeV &\\
  \hline
\end{tabular}\\
}
\caption {Event generation conditions referred to in the text as $cut 1$.}
\label{tab_cuts1}
\end{table}
Events which satisfied these cuts were written in the Les Houches event
format (LHEF~\cite{Boos:2001cv, Alwall:2006yp}) for further processing.

The generators Pythia8 and Herwig++ in the second step were run with
the following non-default configuration.  In order to speed up the simulation
without significantly influencing final results, multiple interactions and
hadronization were turned off in both programs. The QED component of
initial and final state radiation was disabled to avoid double counting of the
radiative corrections, which are calculated in SANC generator in the complete
EW NLO approximation. In Herwig++ less strict than default kinematic
constraints were set for photon momenta: \( k_T(\gamma) > 0.0\) and
\(|\eta_{\gamma}| < 10. \)

In order to avoid edge effects in the distributions after parton showers
were applied the kinematic constraints were strengthened as shown in
table~\ref{tab_cuts2}.
\begin{table}[t]
\centerline{
\begin{tabular}{l|ll}
  \hline
  process & \multicolumn{2}{c}{ cuts} \\
  \hline
  \multirow{2}{*}{$pp\to W^{+}\to l^{+}\nu_{l}(\gamma)+X$} 
      & $M_{inv}(\mu^{+} \nu_{\mu}) > 20$~GeV,
      & $|\eta(\mu^{+}, \nu_{\mu})| < 2.5$, \\
     & $p_T(\mu^{+}, \nu_{\mu}) > 20$~GeV & \\
  \hline
  \multirow{2}{*}{$pp\to Z/\gamma\to l^+l^-(\gamma)+X$ }
      & $M_{inv}(\mu^{+}\mu^{-} ) > 20$~GeV,
      & $|\eta(\mu^{\pm})| < 2.5$, \\
     & $p_T(\mu^{\pm}) > 20$~GeV & \\
  \hline
\end{tabular}\\
}
\caption {Selection criteria applied after showering procedure
  referred to in the text as $cut 2$.}
\label{tab_cuts2}
\end{table}
The lower limit on invariant mass of the leptons was not changed, which would
not lead to edge effects since the transverse momenta constraints for 
$W/Z$ decay products would indirectly increase the actual threshold for
$M_{\ell\ell}$ by a factor of 2.

The physics setup corresponding to the LHC conditions used in this study is specified
in~\cite{Arbuzov:2005dd, Arbuzov:2007db}. The electroweak scheme for the
calculations was chosen to be the $G_{\mu}$-scheme. As parton distribution
functions the MRST2004QED~\cite{Martin:2004dh} set was used since it allows to take
the photon induced contribution into account. The factorization scale was set
to $M_Z$ for neutral current case and to $M_W$ for the charged current.

\subsection{Numerical Results \label{num_sect}}

The results presented in this paragraph were obtained for statistics of $7
\times 10^7$ events for each channel (CC and NC) calculated in both LO and EW
NLO approximation.  The data produced in the wide selection criteria ($cut 1$)
were then subjected to the selection ($cut 2$) with $\sim 50\%$ efficiency. 
\begin{table}[t]
\centerline{
\begin{tabular}{|l|c|c|c|c|c|c|}
\hline
& \multicolumn{3}{|c|}{$NC$} & \multicolumn{3}{|c|}{$CC$} \\
& $\sigma_{lo} [pb]$ & $\sigma_{nlo} [pb]$&$\delta [\%]$&$\sigma_{lo} [pb]$&$\sigma_{nlo} [pb]$&$\delta [\%]$ \\
\hline
$\mathrm{HP}_{\tt SANC}, cut 1$ 		& 2332(1) & 2390(1)& 2.5(1) & 9764(1) & 9729(1) & -0.36(1)\\
$\mathrm{HP}_{\tt SANC}, cut 2$ 		& 807.7(4)   & 785.6(4) & -2.7(1) & 5428(1) & 5296(1) & -2.4(1)\\
$\mathrm{HP}_{\tt SANC}+\mathrm{PS}_{\tt Herwig++}, cut 2$& 771.3(4) & 752.0(4) & -2.5(1) & 4917(1) & 4807(1) & -2.2(1)\\
$\mathrm{HP}_{\tt SANC}+\mathrm{PS}_{\tt Pythia8}, cut 2$	& 785.8(4) & 762.3(4) & -3.0(1) & 5033(1) & 4919(1) & -2.3(1)\\
\hline
\end{tabular}
}
\caption {Inclusive cross sections and relative EW corrections.}
\label{tab_csec_cuts}
\end{table}
Table \ref{tab_csec_cuts} shows the effect of this selection on the inclusive cross
section and electroweak NLO correction values.  Here $\delta$ denotes the
relative corrections, $\delta = (\sigma_{NLO}/\sigma_{LO} - 1) \times 100\%$.
Expressions like ``$\mathrm{HP}_{\tt SANC}+\mathrm{PS}_{\tt Herwig++}$'' denote
the case when the hard process (HP) data were generated with the SANC generator and
then processed with Herwig++ to apply parton showers (PS). The first and
second rows in the table show the generator-level cross sections calculated
with the SANC generator before parton shower algorithms applied in the $cut
1$ and $cut 2$ conditions, respectively. The third and fourth rows show 
effects of the $cut 2$ selection when parton showers via Herwig++ and
Pythia8 were applied.

To compare the parton shower algorithms in Pythia8 and Herwig++ it is
convenient to introduce a parameter
\begin{align*}
R_{X} = \frac{d\sigma_{\tt Herwig++}/dX}{d\sigma_{\tt Pythia8}/dX},
\end{align*}
where $d\sigma_{\tt Pythia8}/dX$ represents a differential cross section by an
observable $X$ calculated with parton showers applied by Pythia8.

\begin{figure}[!h]
\begin{center}
\subfloat{
  \includegraphics[width=6.0cm, height=5.0cm]{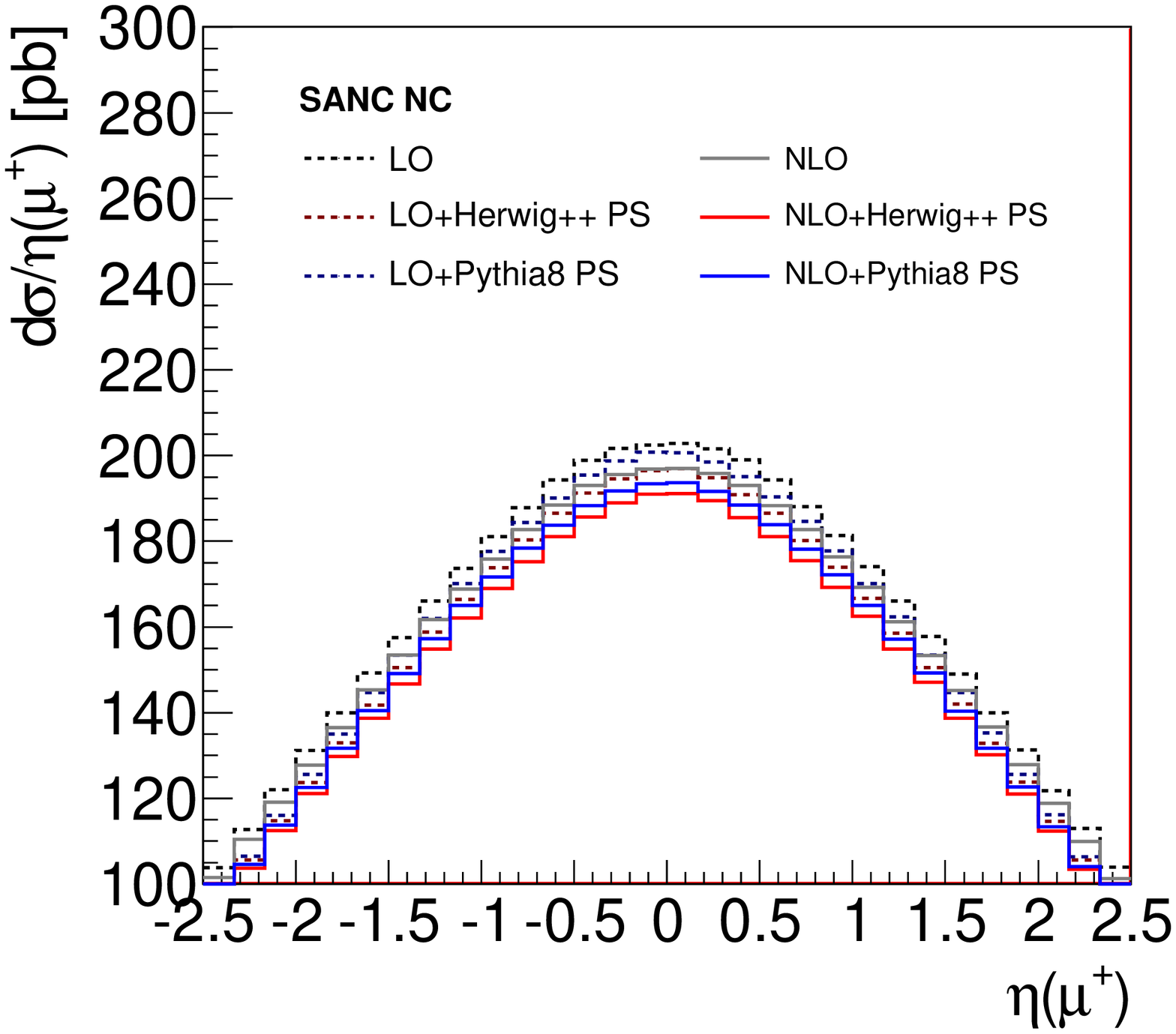}
  \includegraphics[width=6.0cm, height=5.0cm]{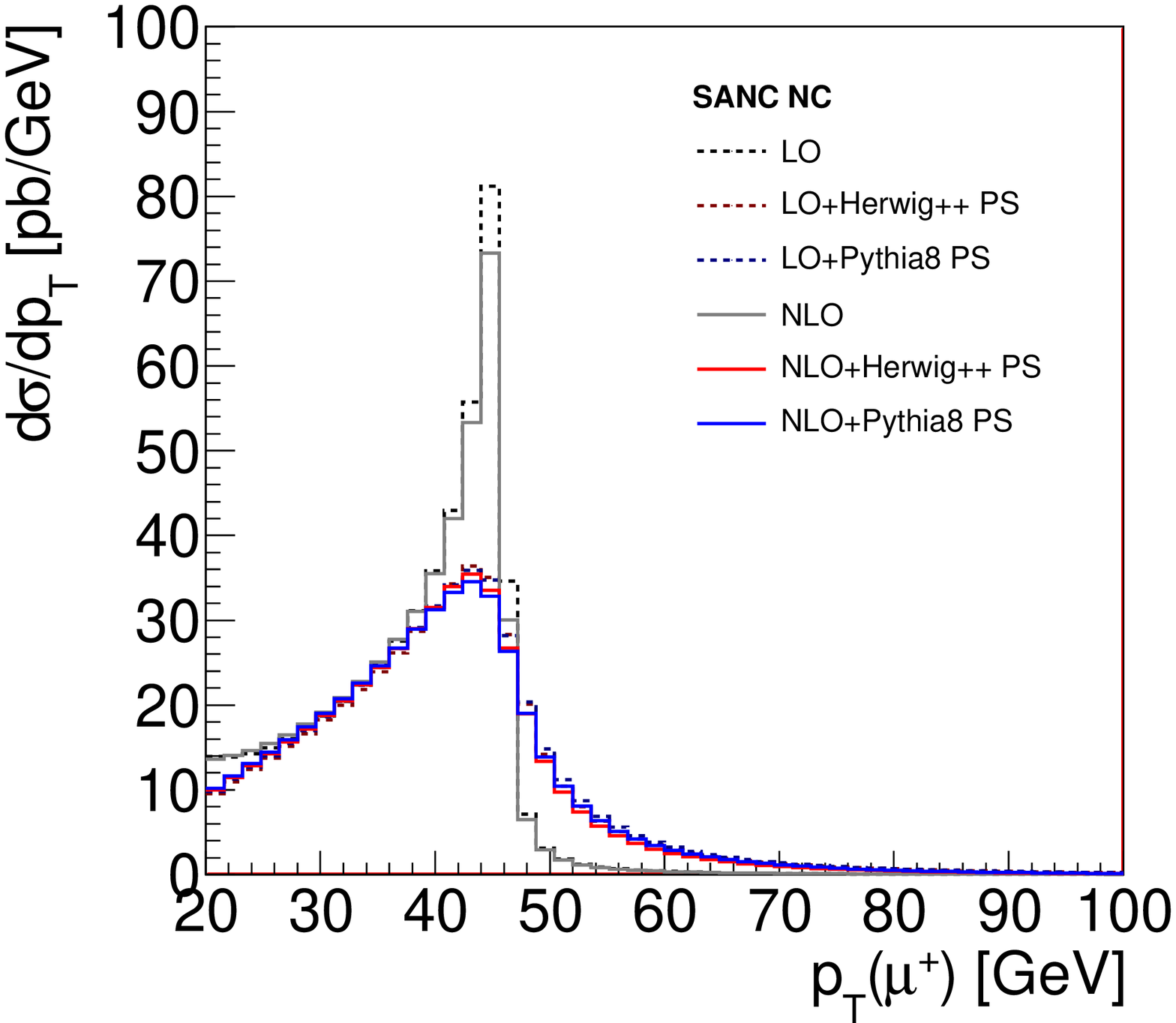}
}\\
\subfloat{
  \includegraphics[width=6.0cm, height=5.0cm]{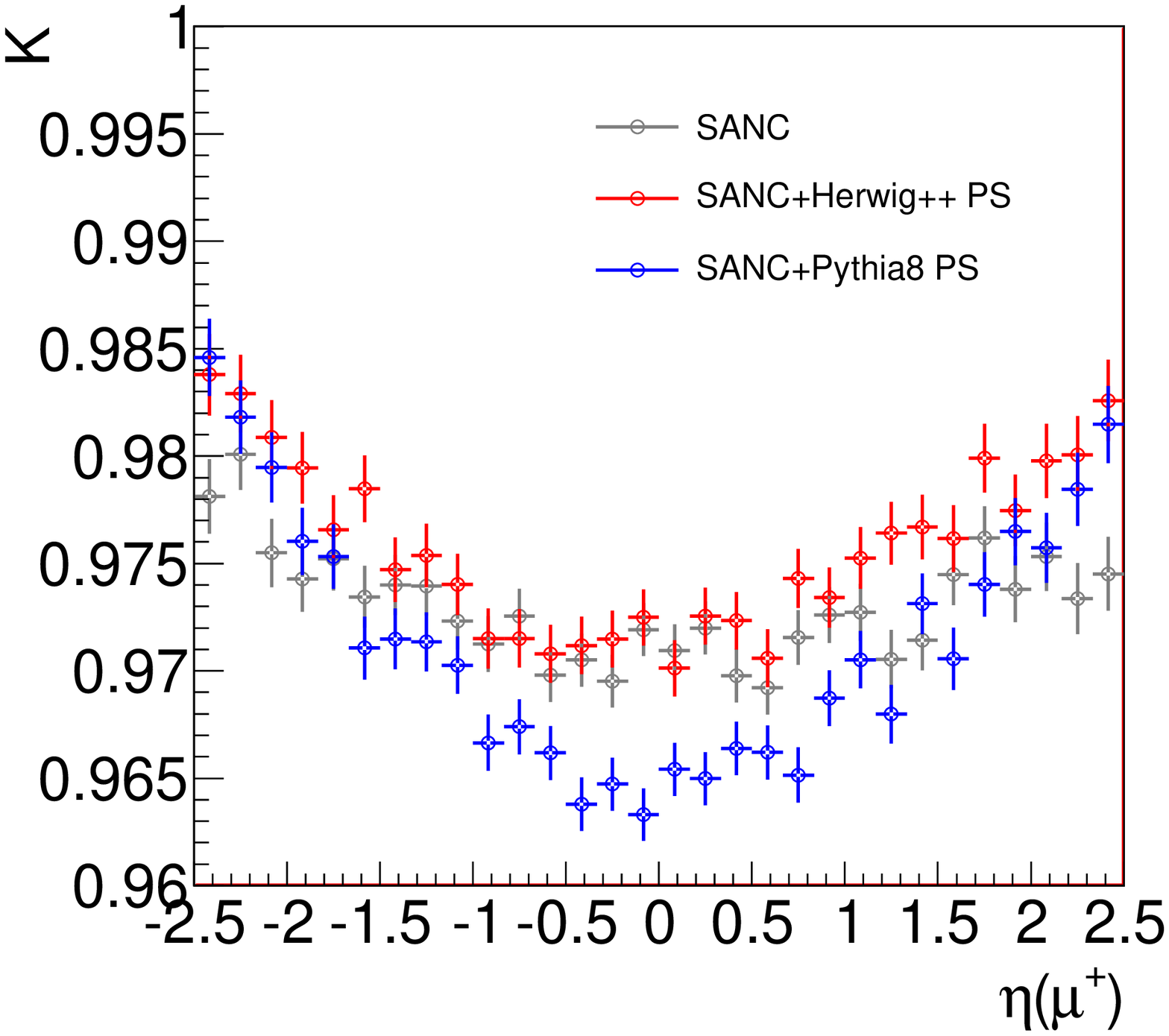}
  \includegraphics[width=6.0cm, height=5.0cm]{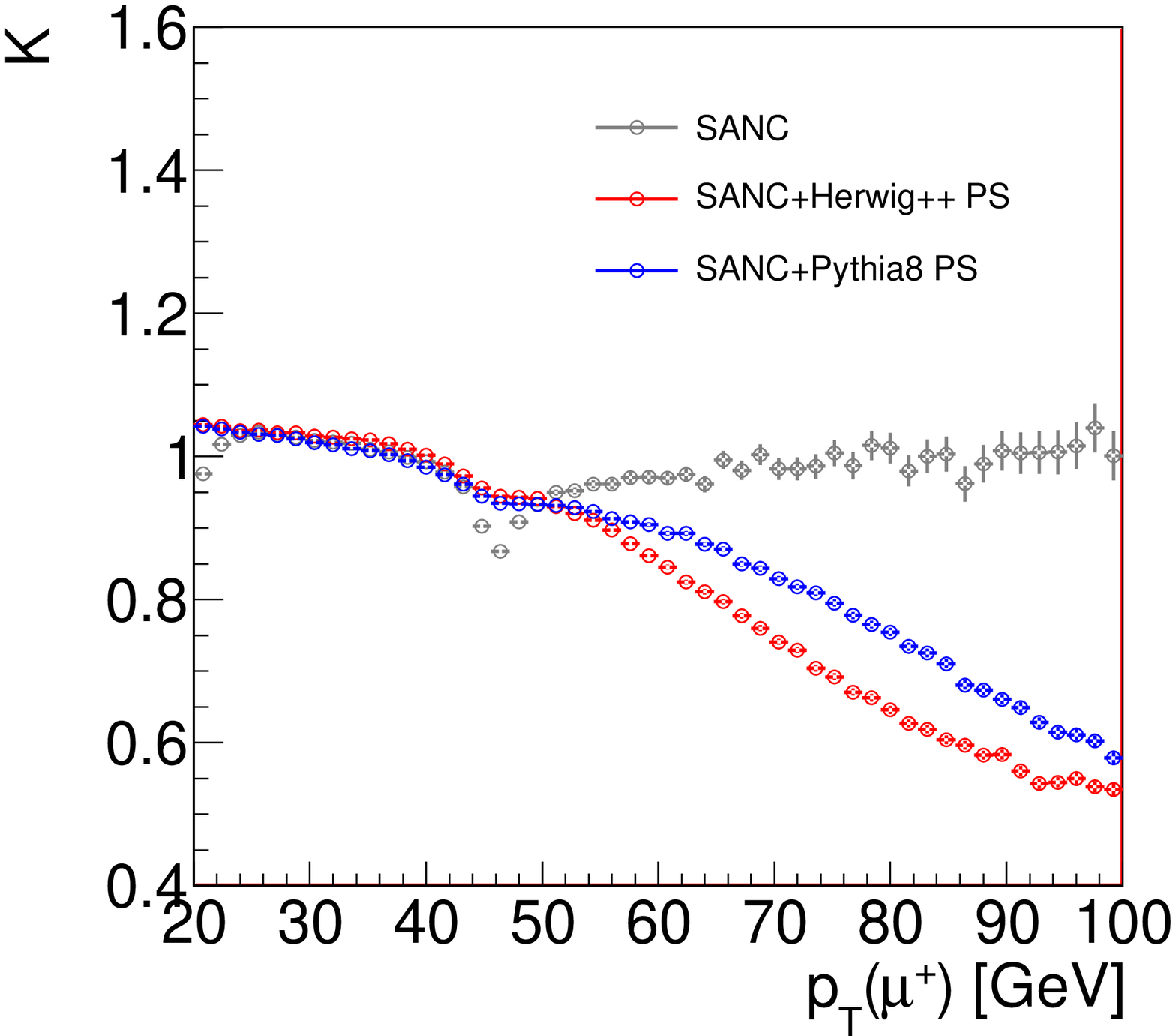}
}\\
\subfloat{
  \includegraphics[width=6.0cm, height=5.0cm]{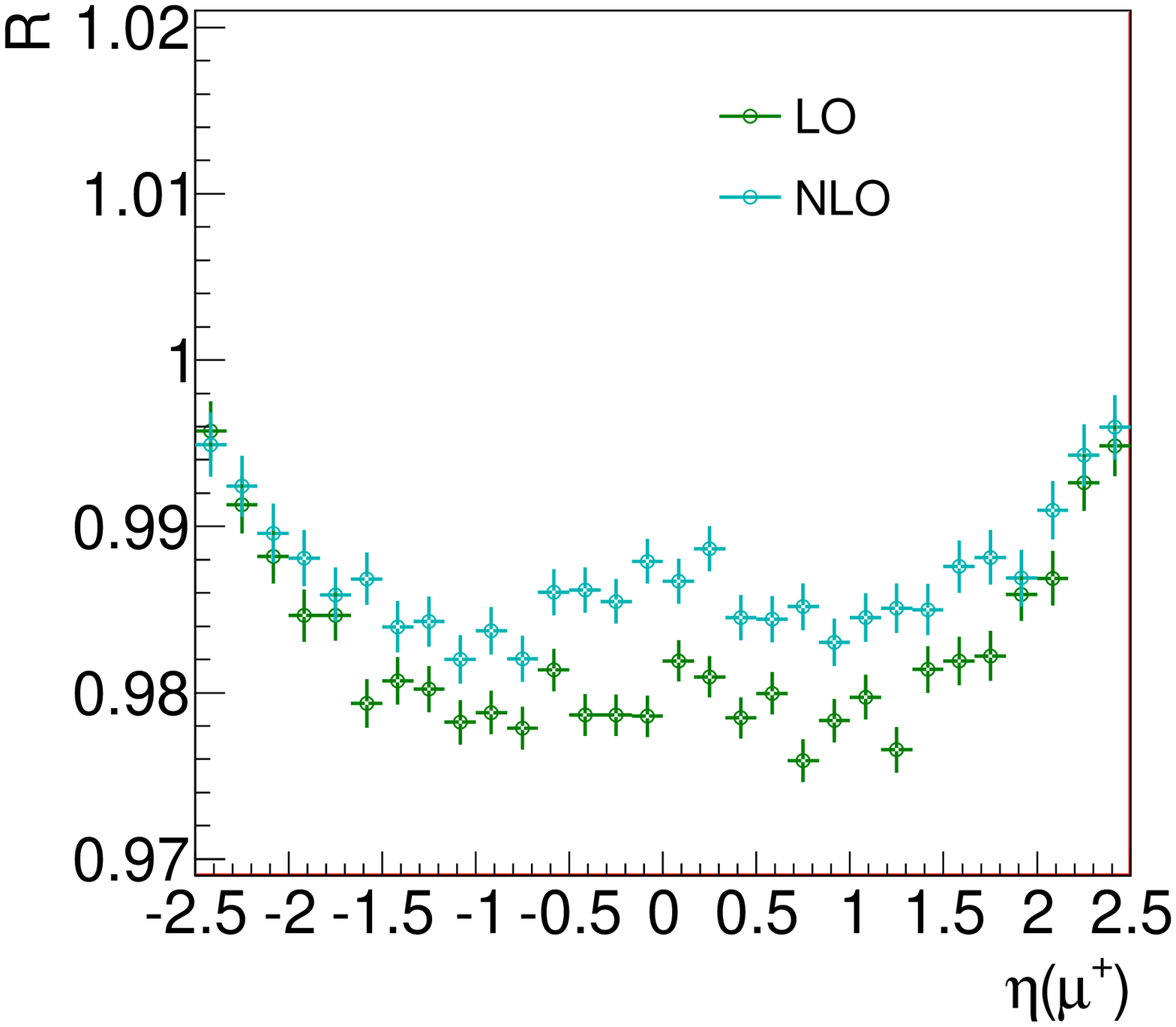}
  \includegraphics[width=6.0cm, height=5.0cm]{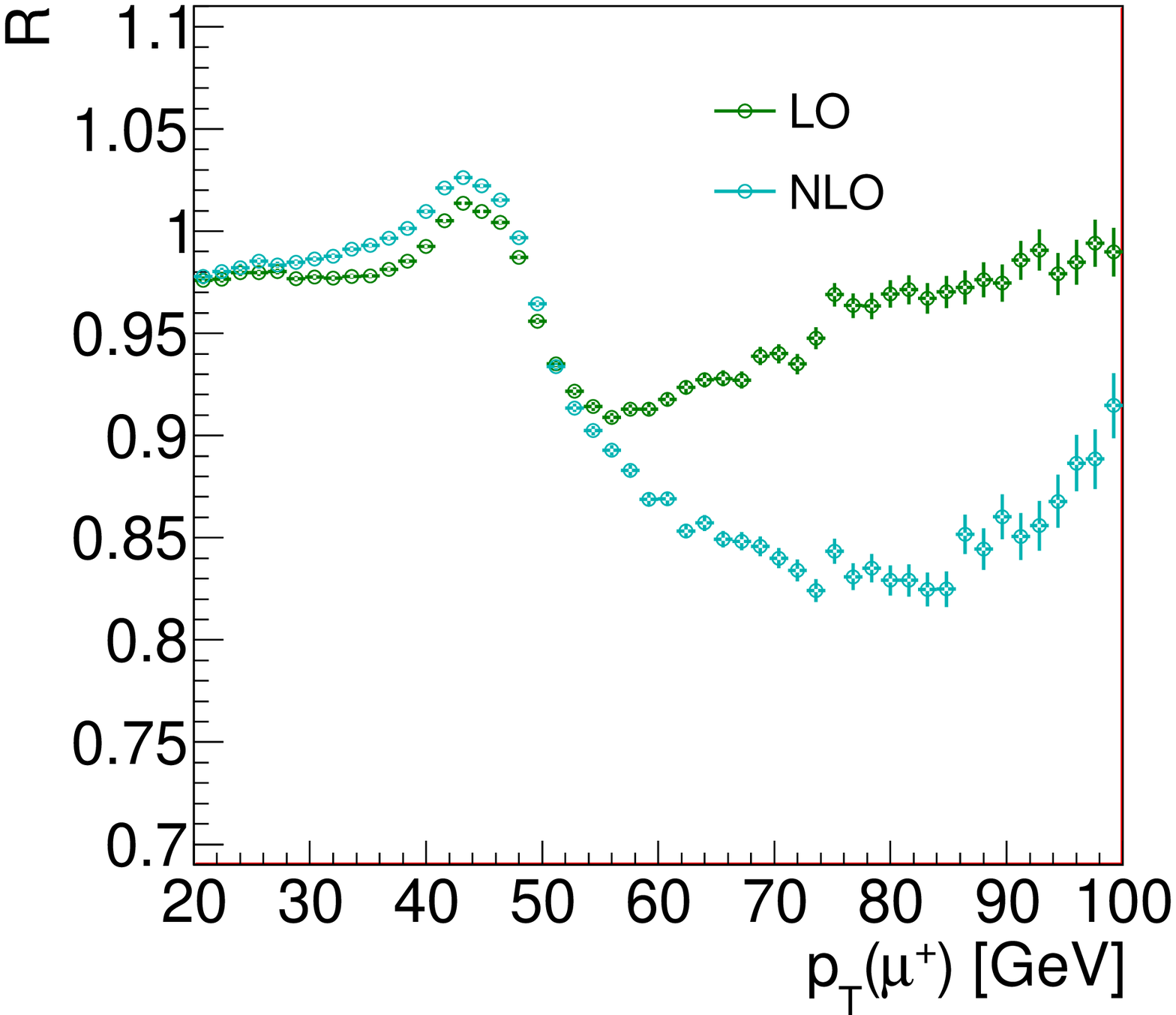}
}
\end{center}
\caption {Distributions of pseudorapidity (\textit{left}) and transverse momentum 
(\textit{right}) of $\mu^{+}$ for NC DY.}
\label{fig_nc_mu}
\end{figure}

\begin{figure}[t]
\begin{center}
\subfloat{
\includegraphics[width=6.0cm, height=5.0cm]{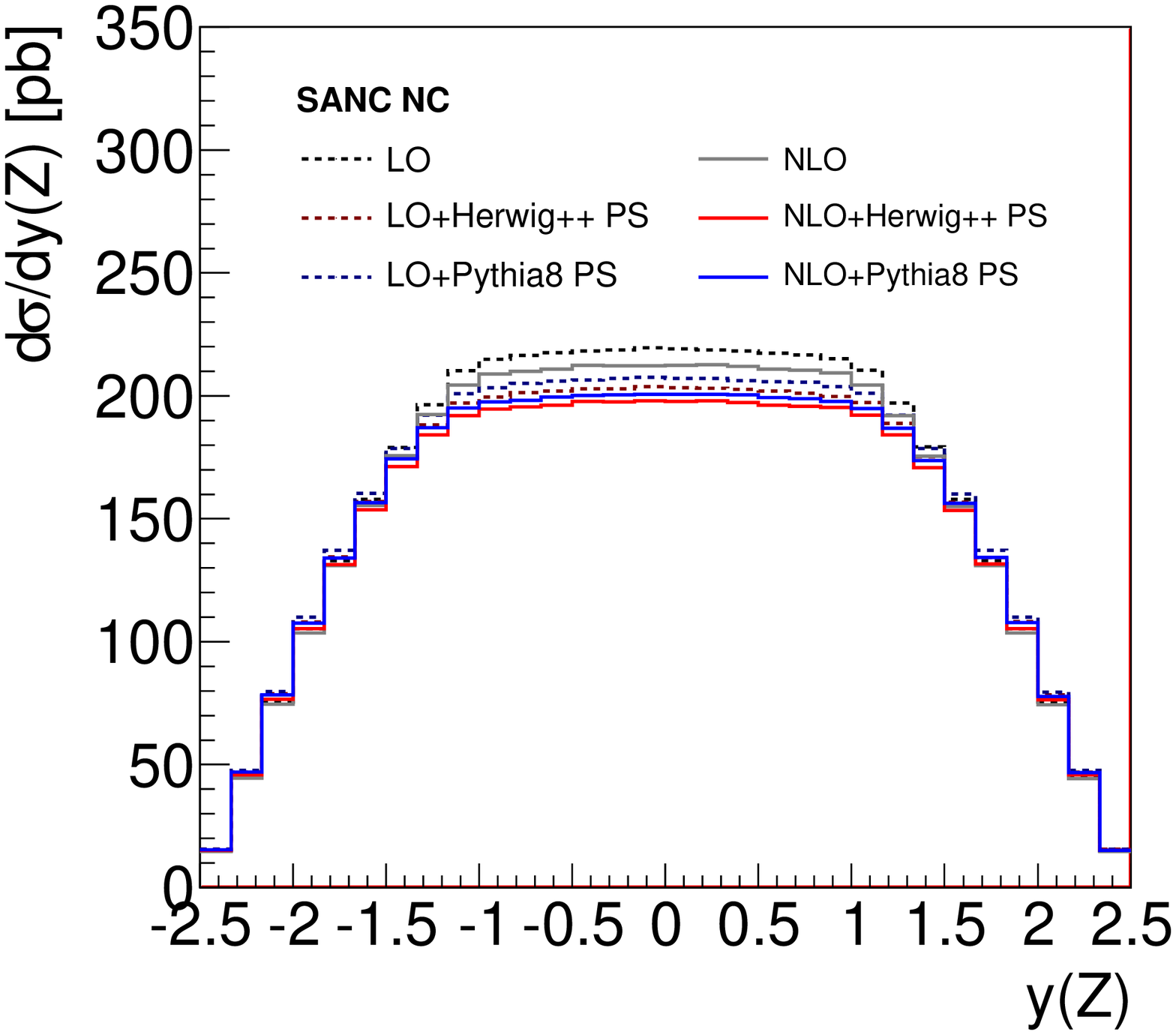}
\includegraphics[width=6.0cm, height=5.0cm]{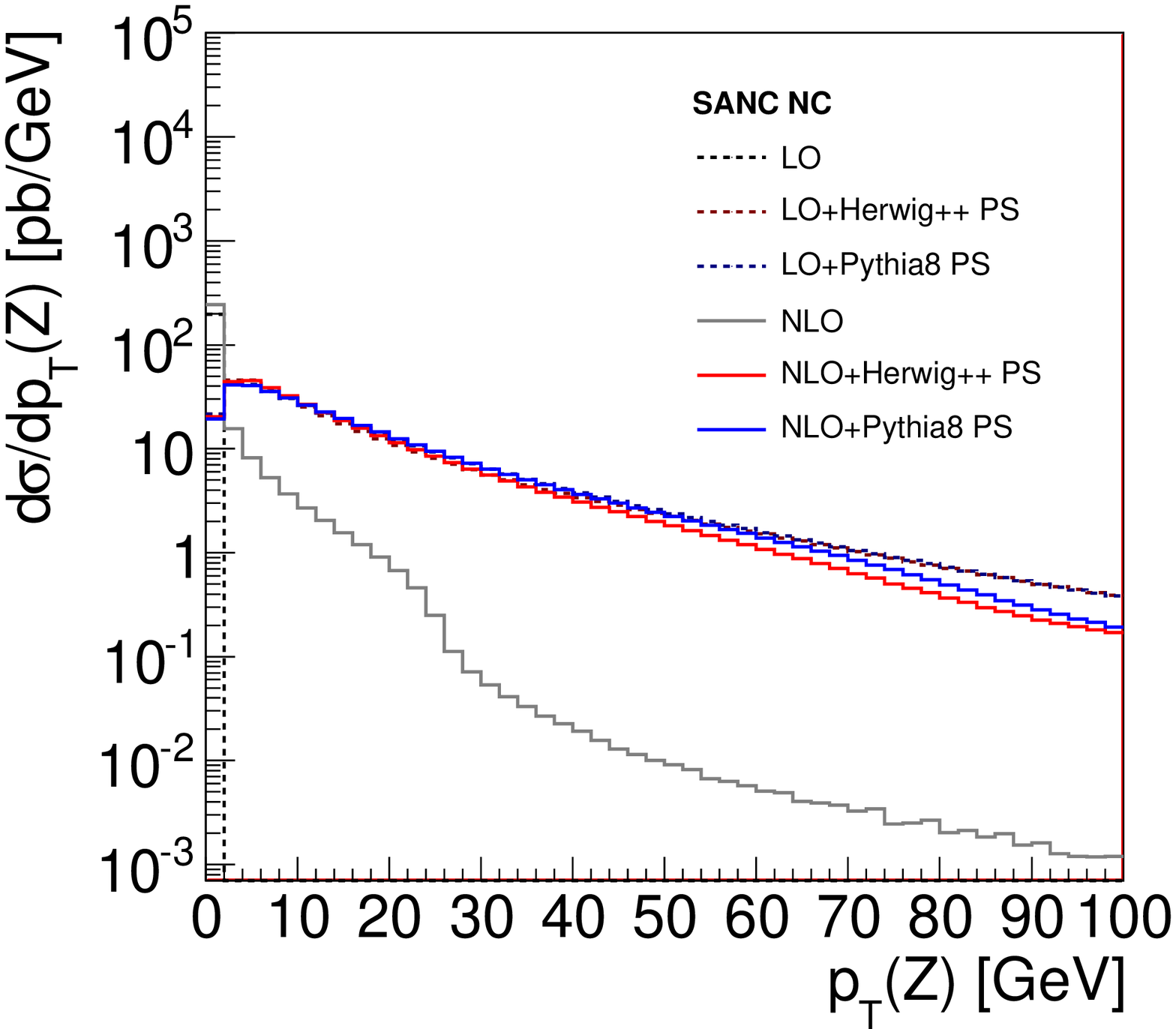}}\\
\subfloat{
\includegraphics[width=6.0cm, height=5.0cm]{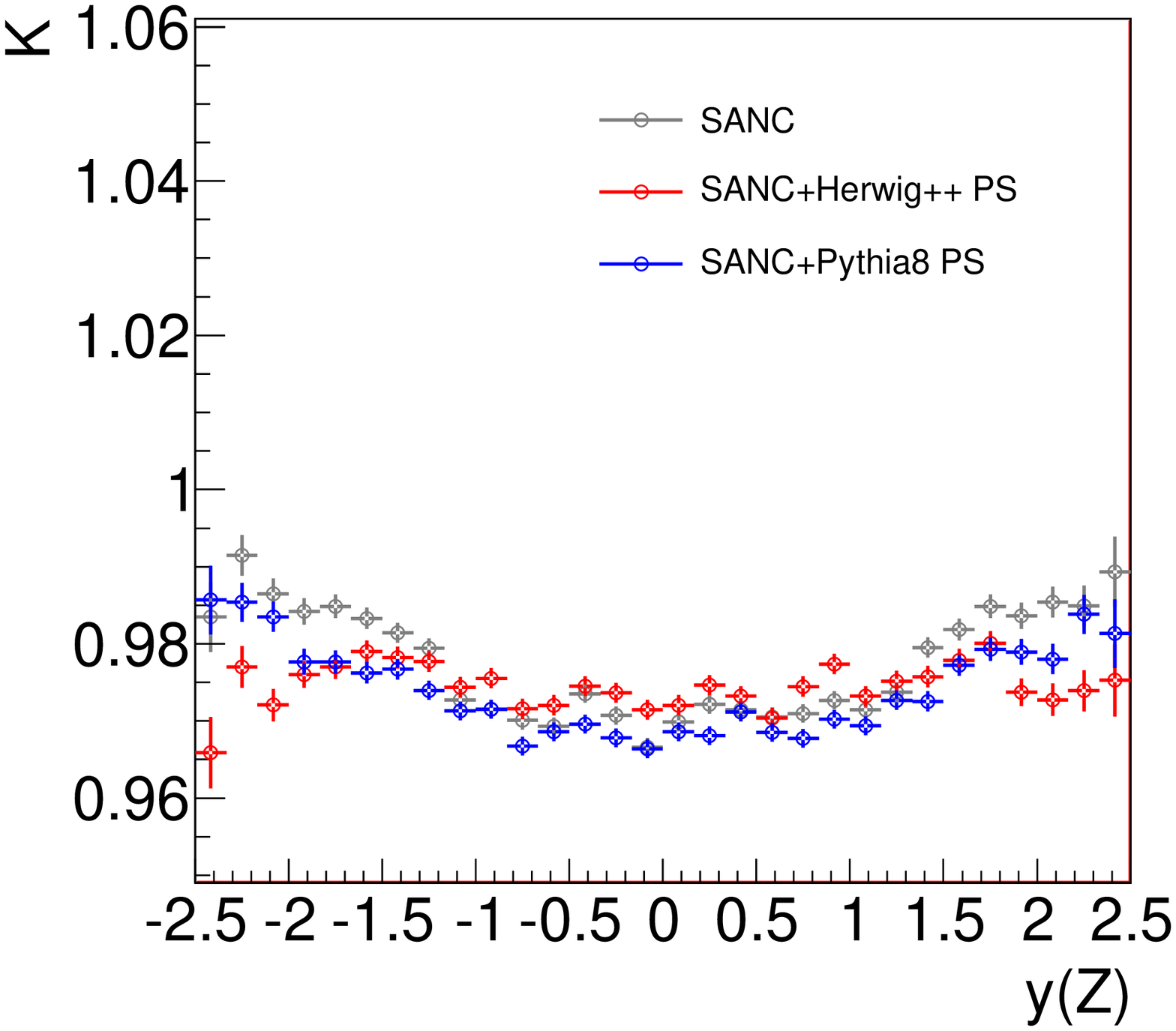}
\includegraphics[width=6.0cm, height=5.0cm]{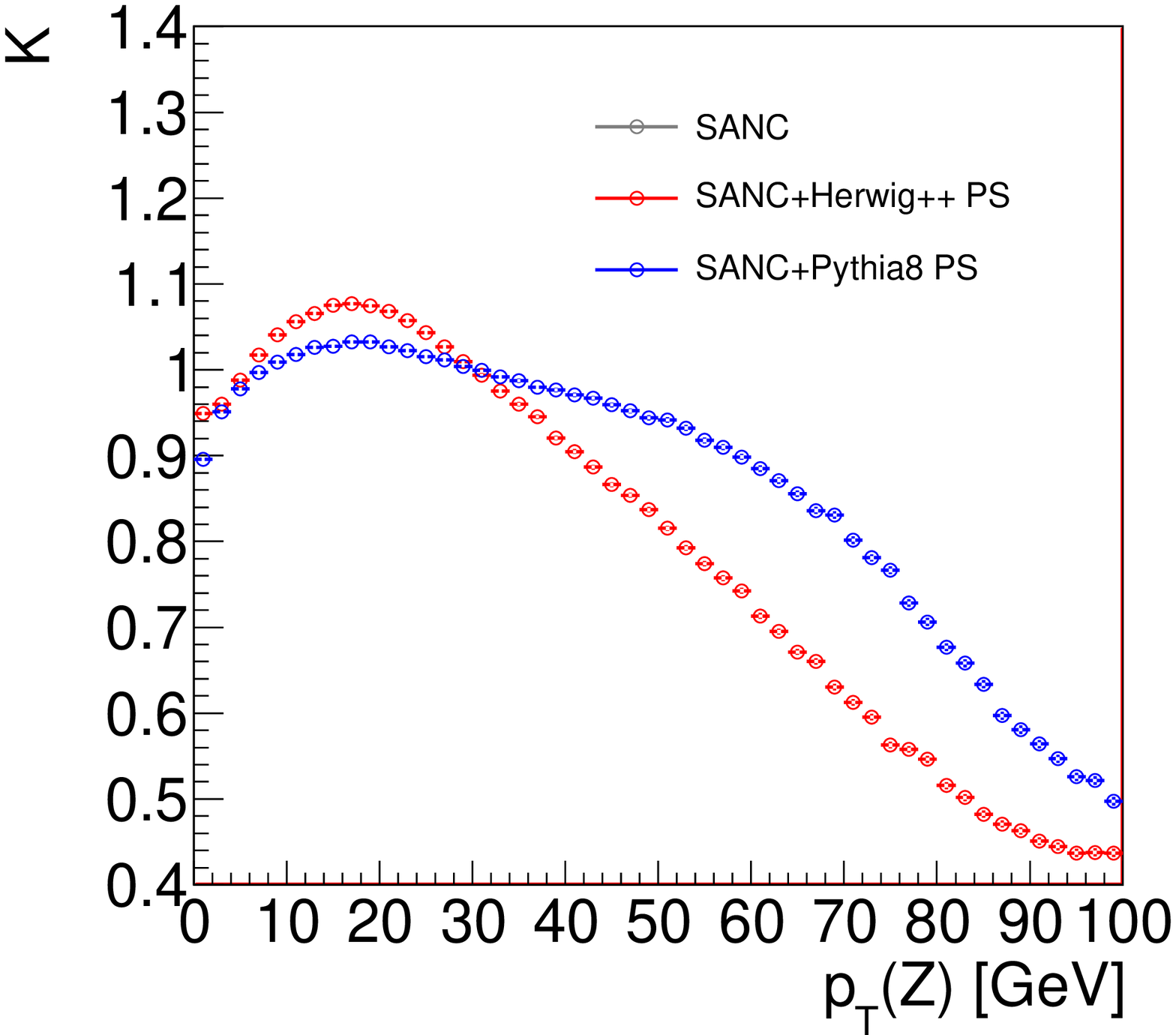}}\\
\subfloat{
\includegraphics[width=6.0cm, height=5.0cm]{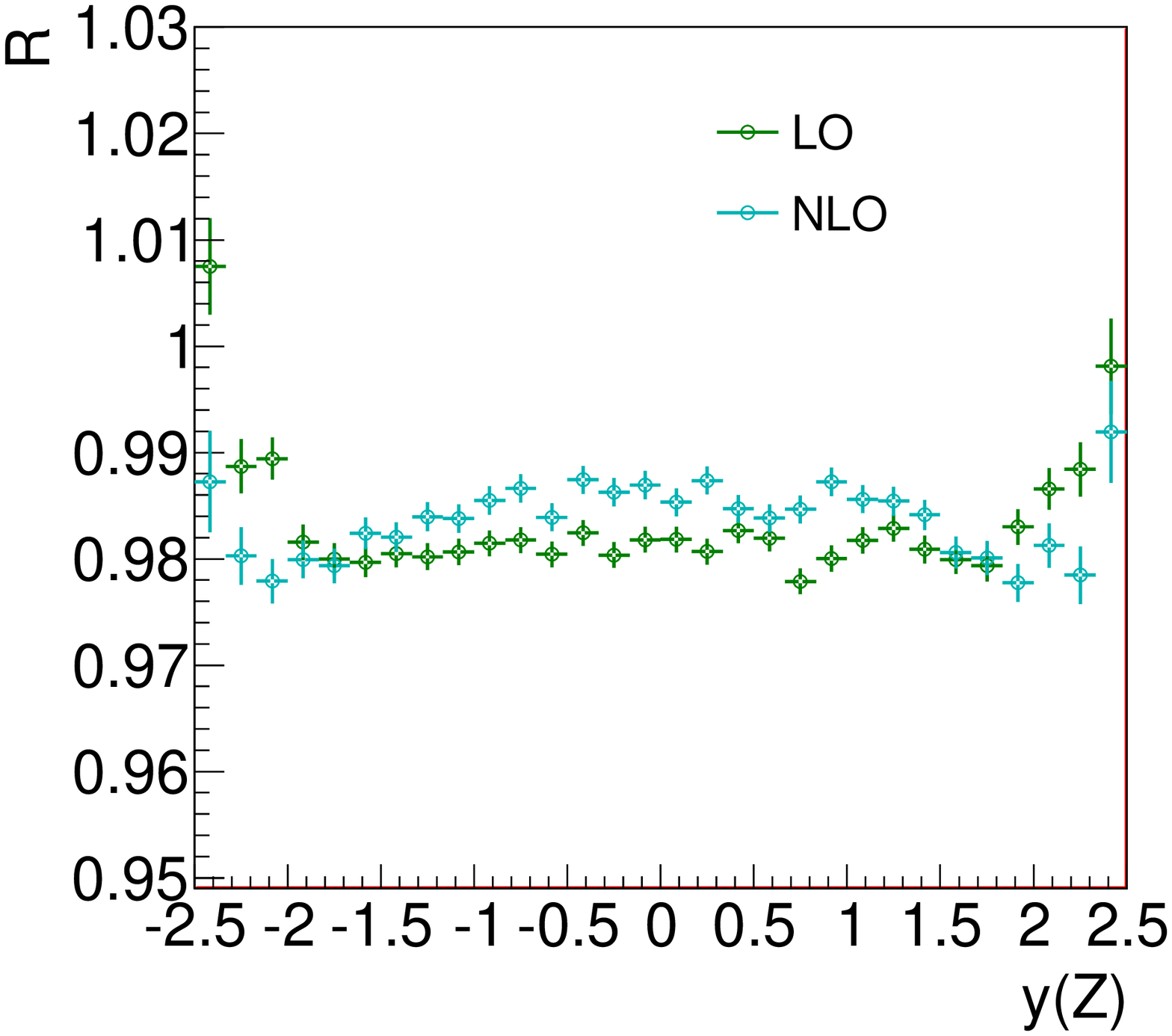}
\includegraphics[width=6.0cm, height=5.0cm]{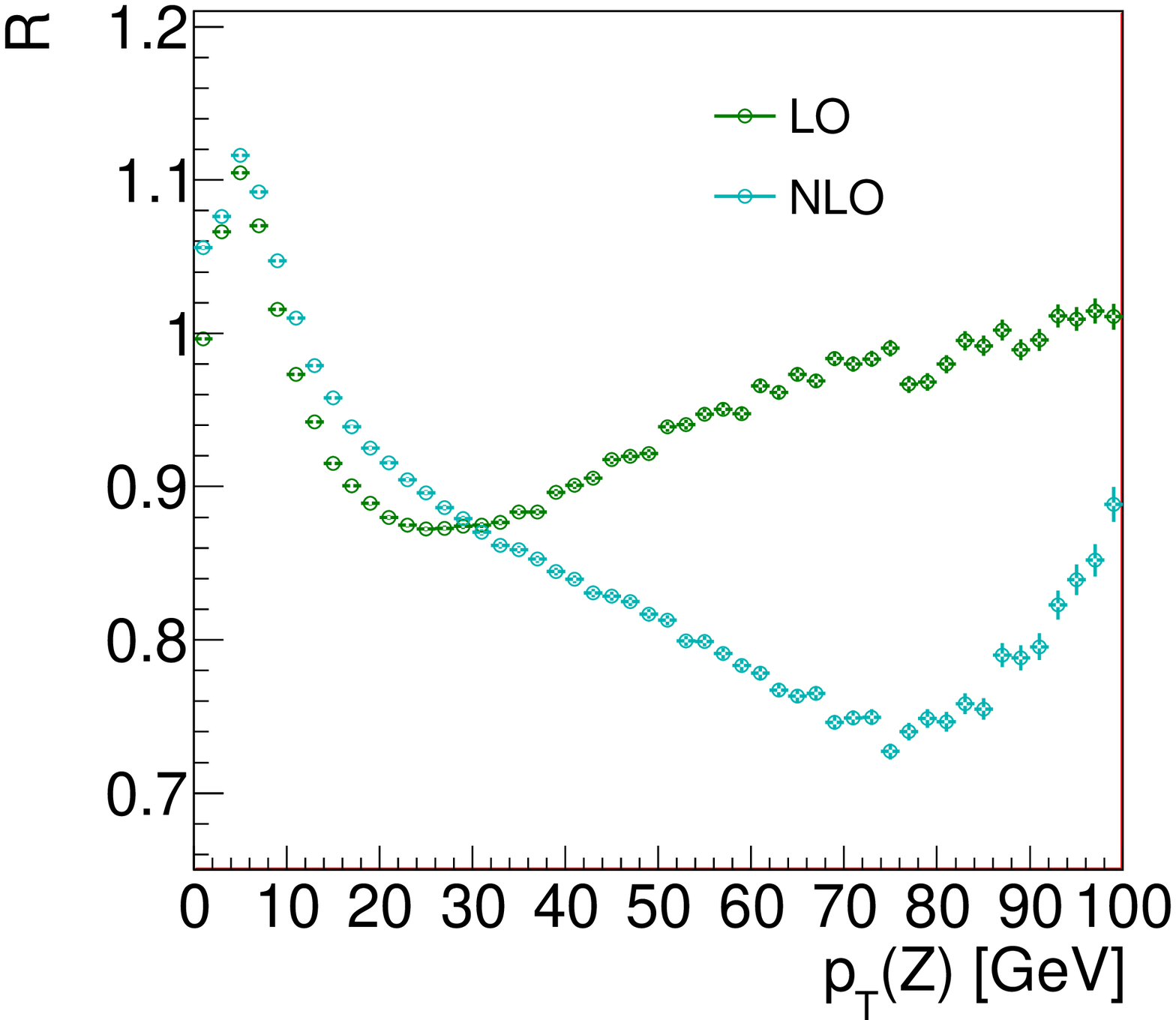}}
\end{center}
\caption {Distributions of rapidity (\textit{left}) and transverse momentum 
(\textit{right}) of $Z$ boson for NC DY.}
\label{fig_nc_Z}
\end{figure}
\begin{figure}[t]
\begin{center}
\includegraphics[width=6.0cm, height=5.0cm]{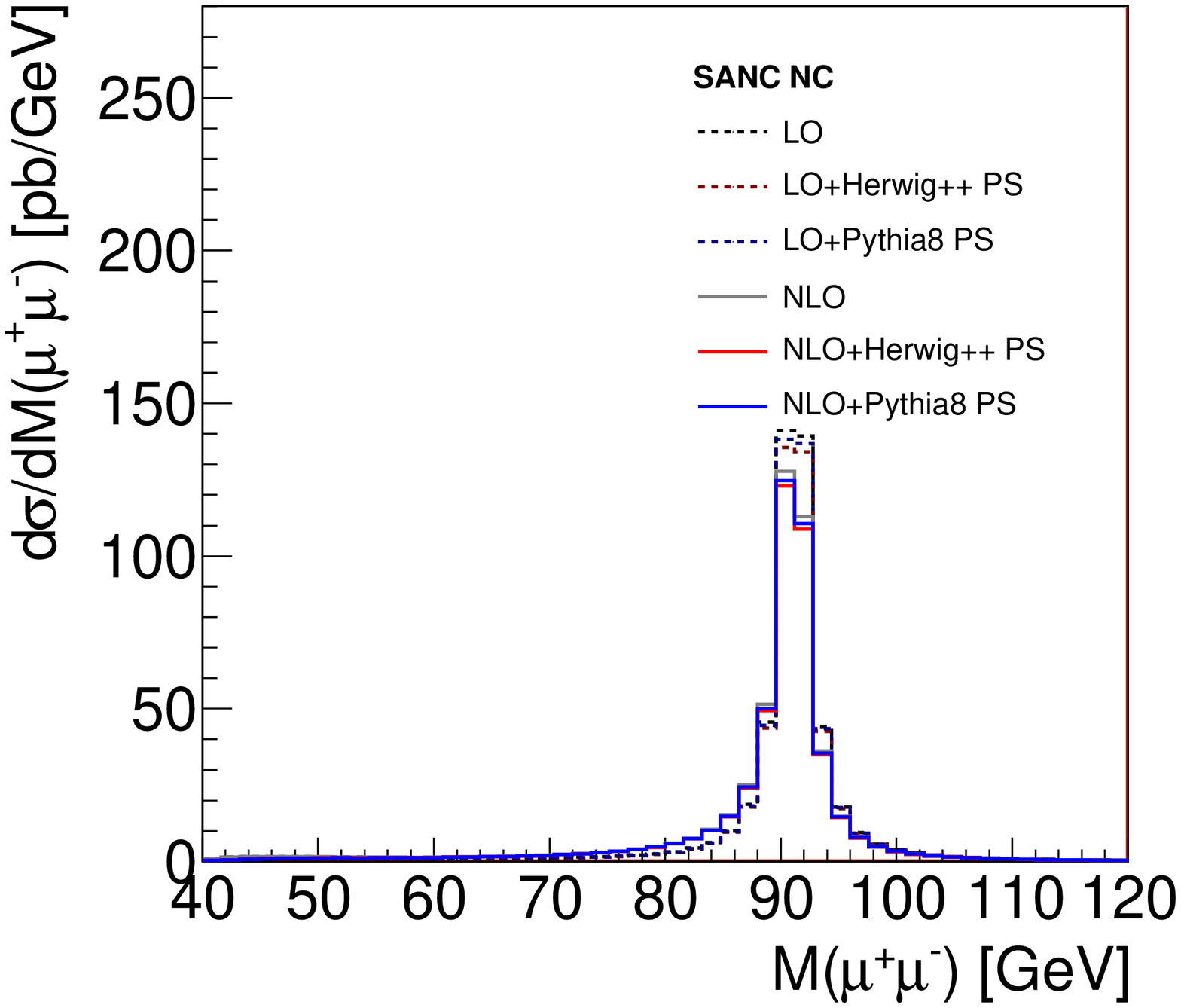}\\
\includegraphics[width=6.0cm, height=5.0cm]{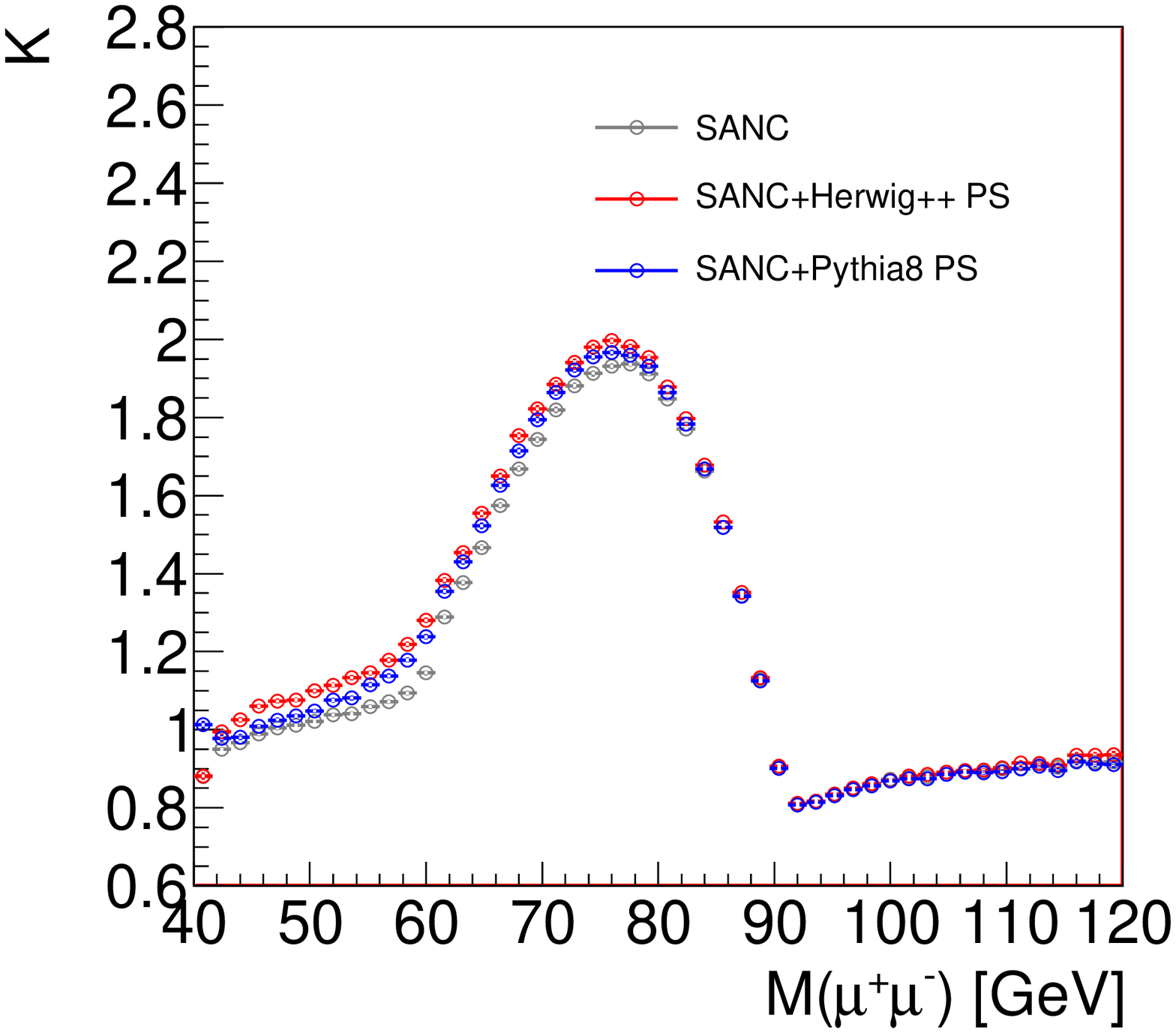}\\
\includegraphics[width=6.0cm, height=5.0cm]{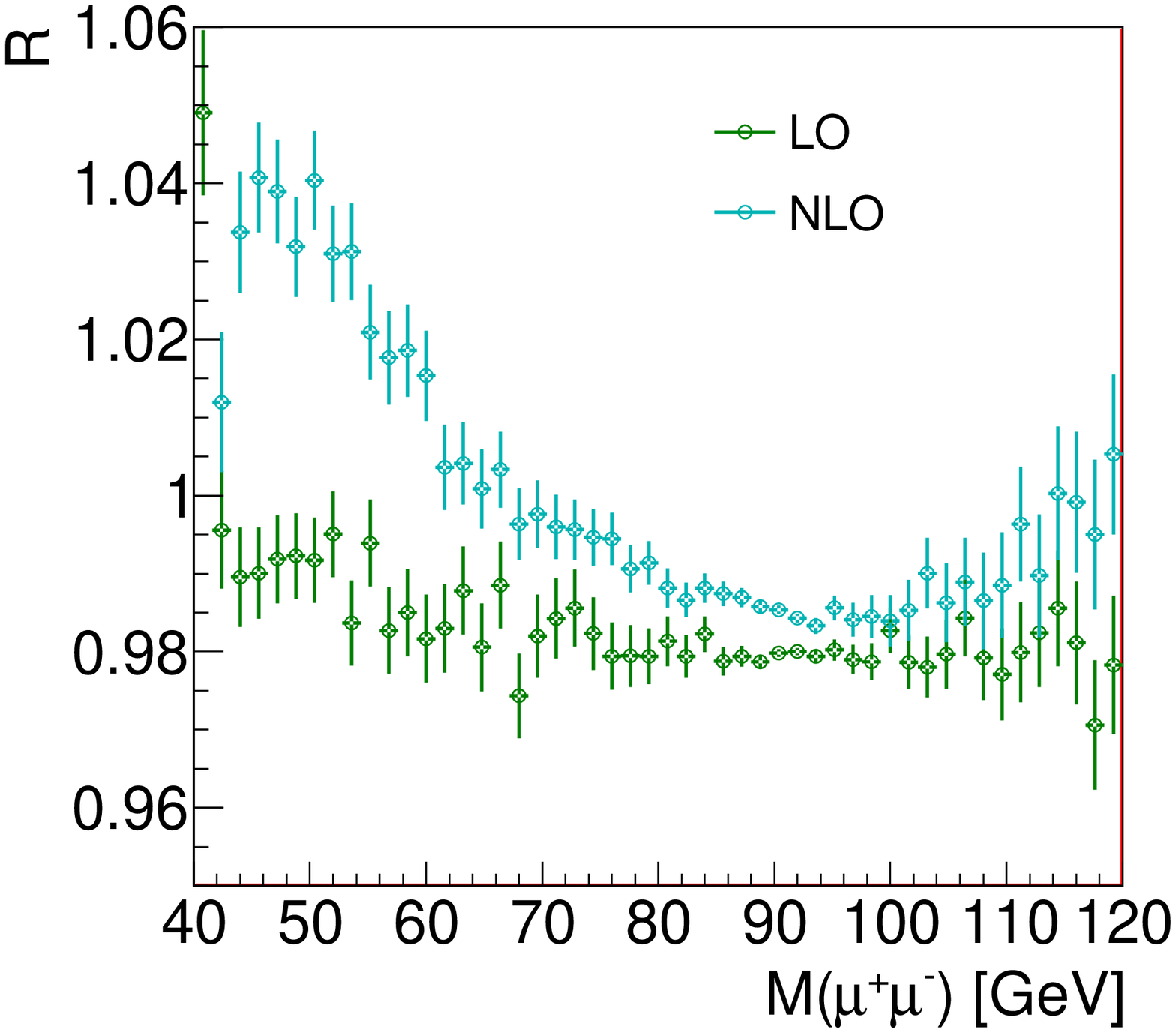}
\end{center}
\caption {Distributions of invariant mass of $\mu^+, \mu^-$ pair for NC DY.}
\label{fig_nc_minvZ}
\end{figure}
\begin{figure}
\begin{center}
\subfloat{
\includegraphics[width=6.0cm, height=5.0cm]{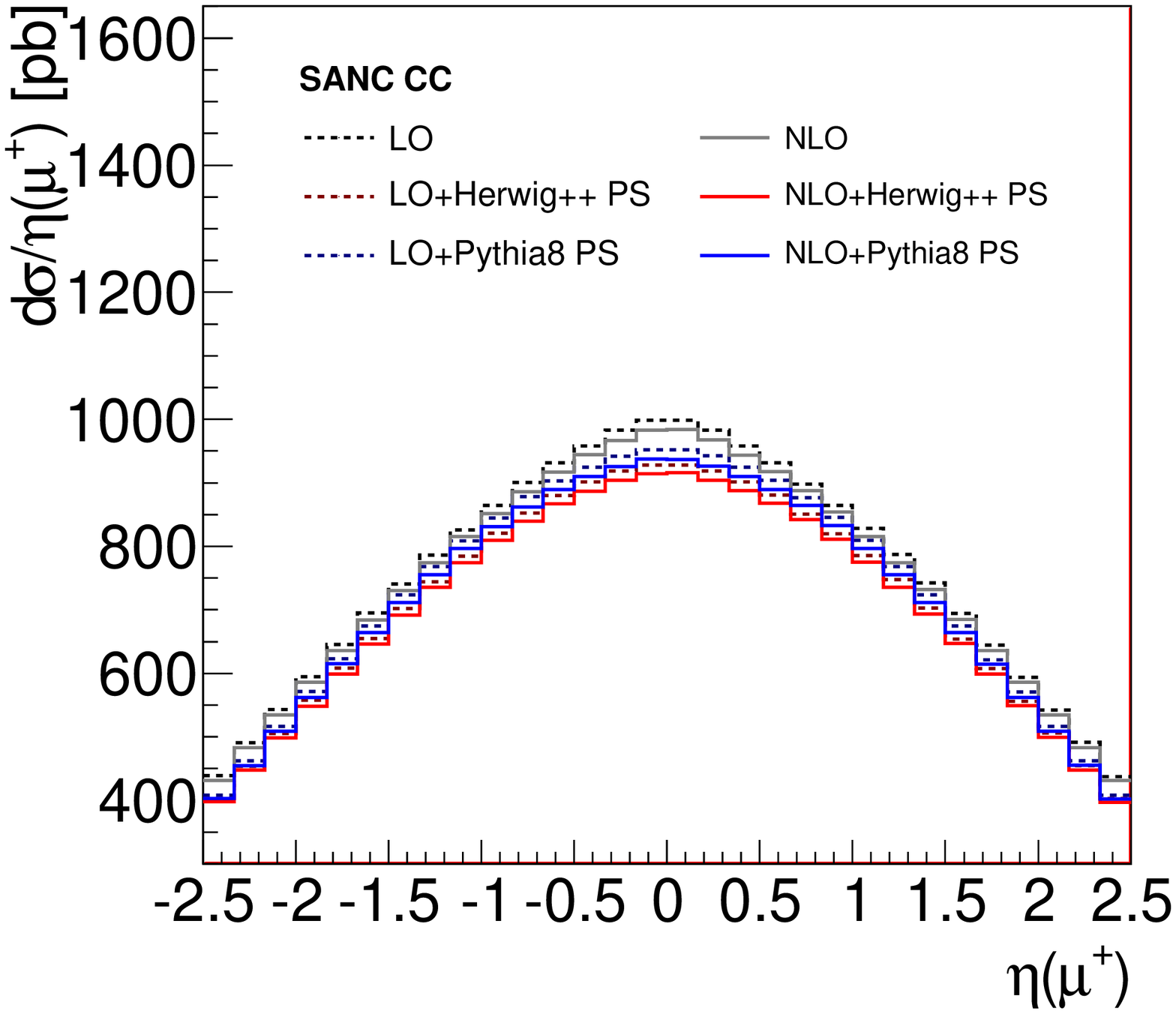}
\includegraphics[width=6.0cm, height=5.0cm]{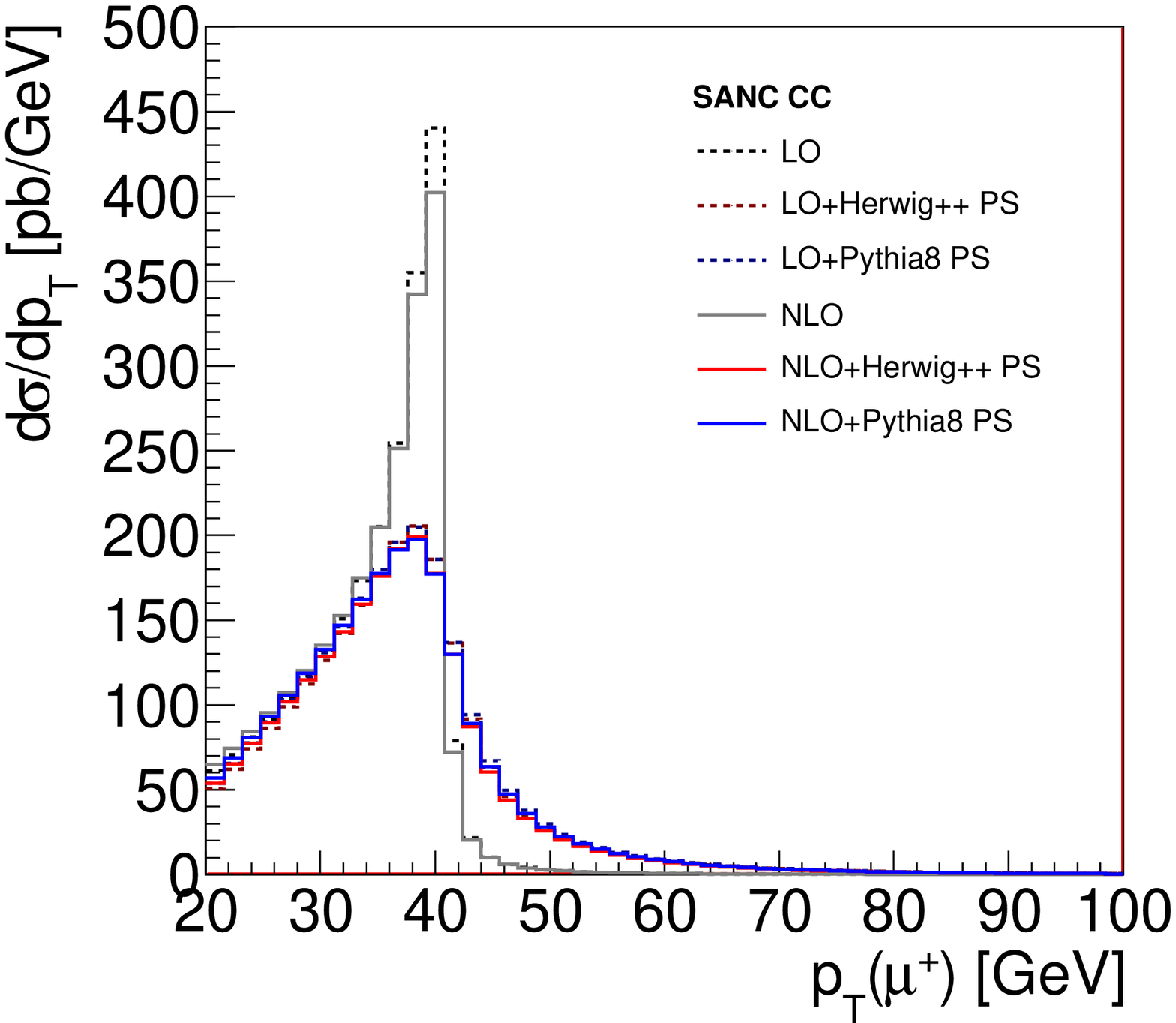}}\\
\subfloat{
\includegraphics[width=6.0cm, height=5.0cm]{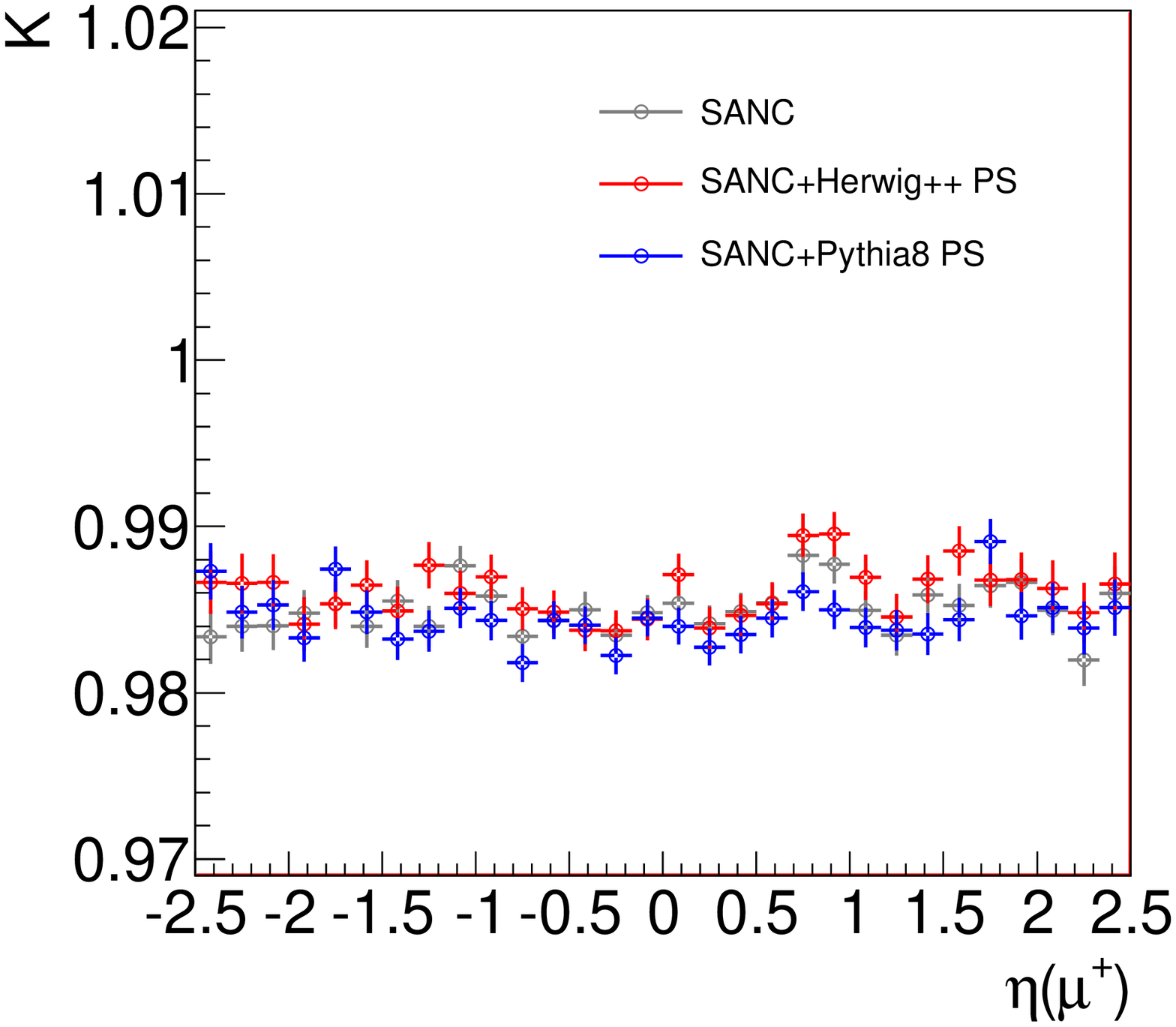}
\includegraphics[width=6.0cm, height=5.0cm]{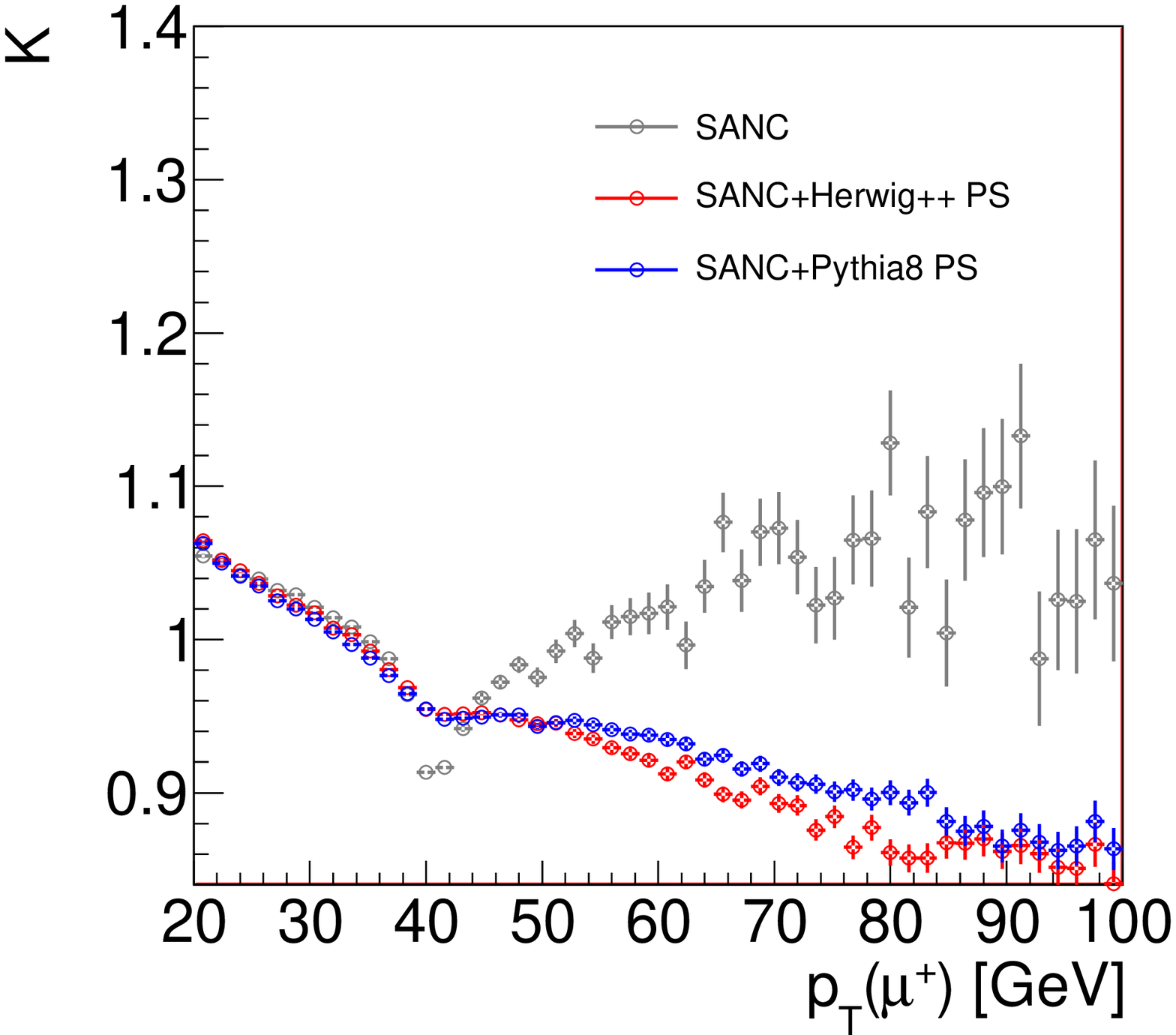}}\\
\subfloat{
\includegraphics[width=6.0cm, height=5.0cm]{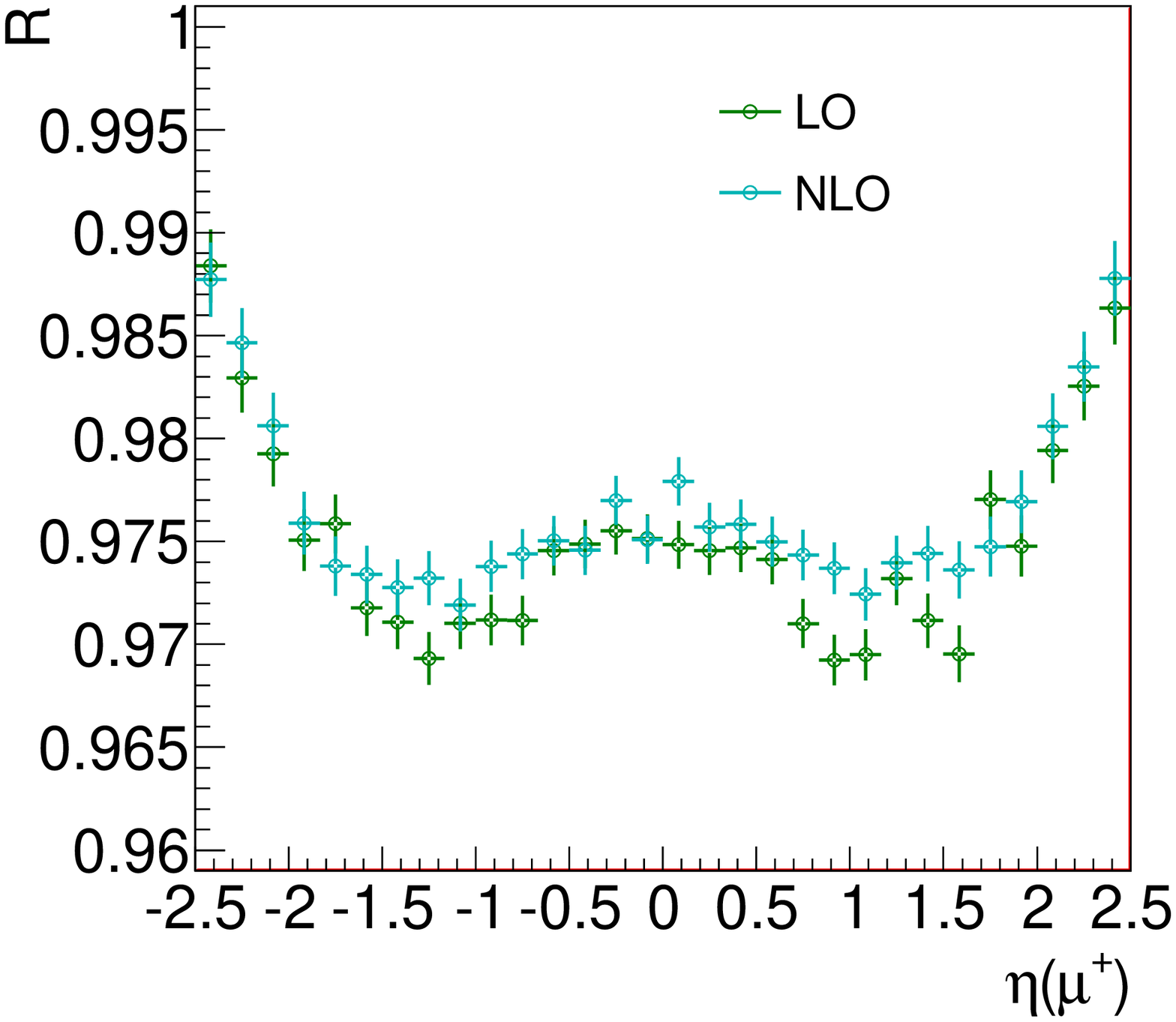}
\includegraphics[width=6.0cm, height=5.0cm]{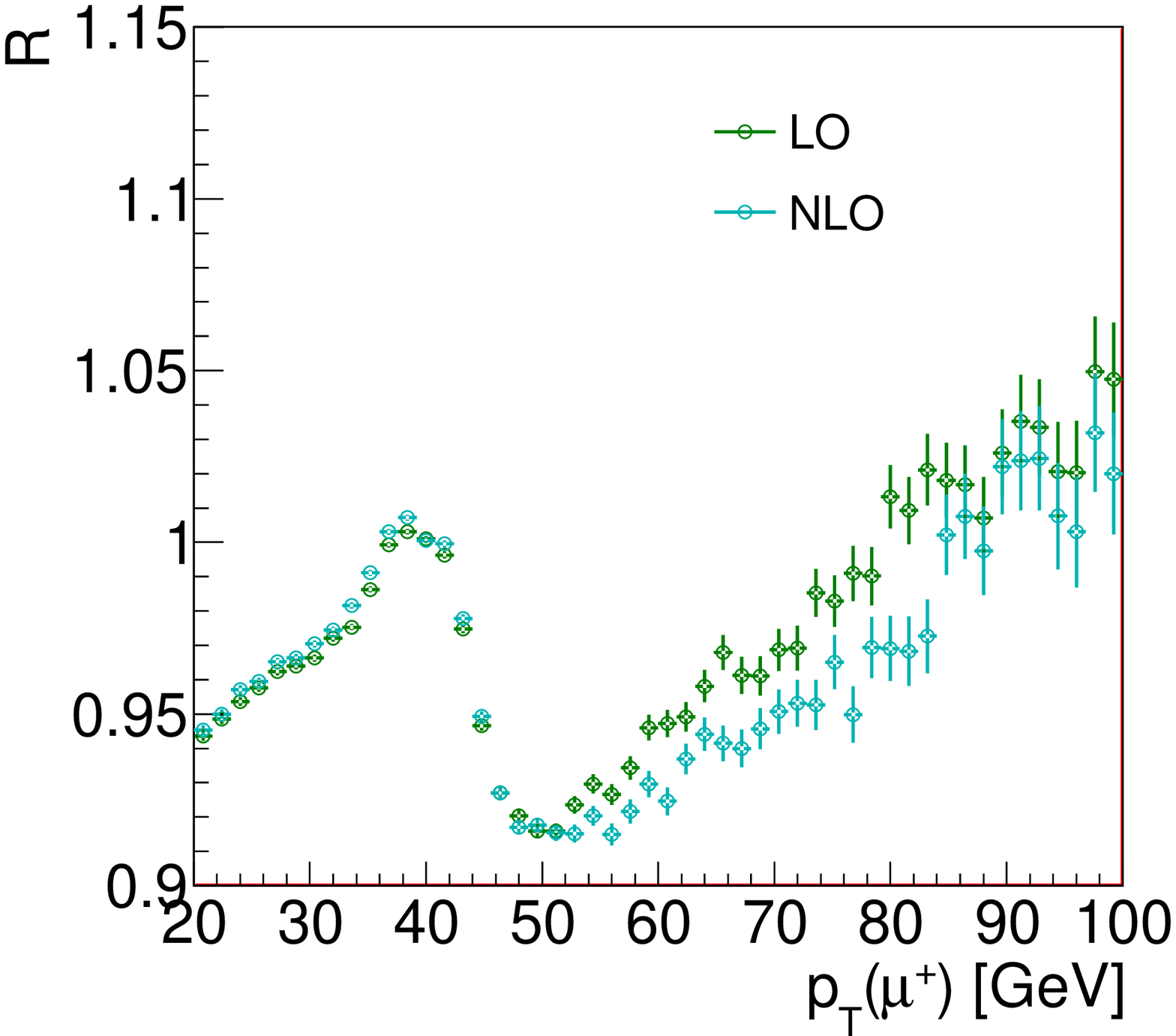}}
\end{center}
\caption {Distributions of pseudorapidity (\textit{left}) and transverse momentum 
(\textit{right}) of $\mu^{+}$ for CC DY.}
\label{fig_cc_mu}
\end{figure}
\begin{figure}
\begin{center}
\subfloat{
\includegraphics[width=6.0cm, height=5.0cm]{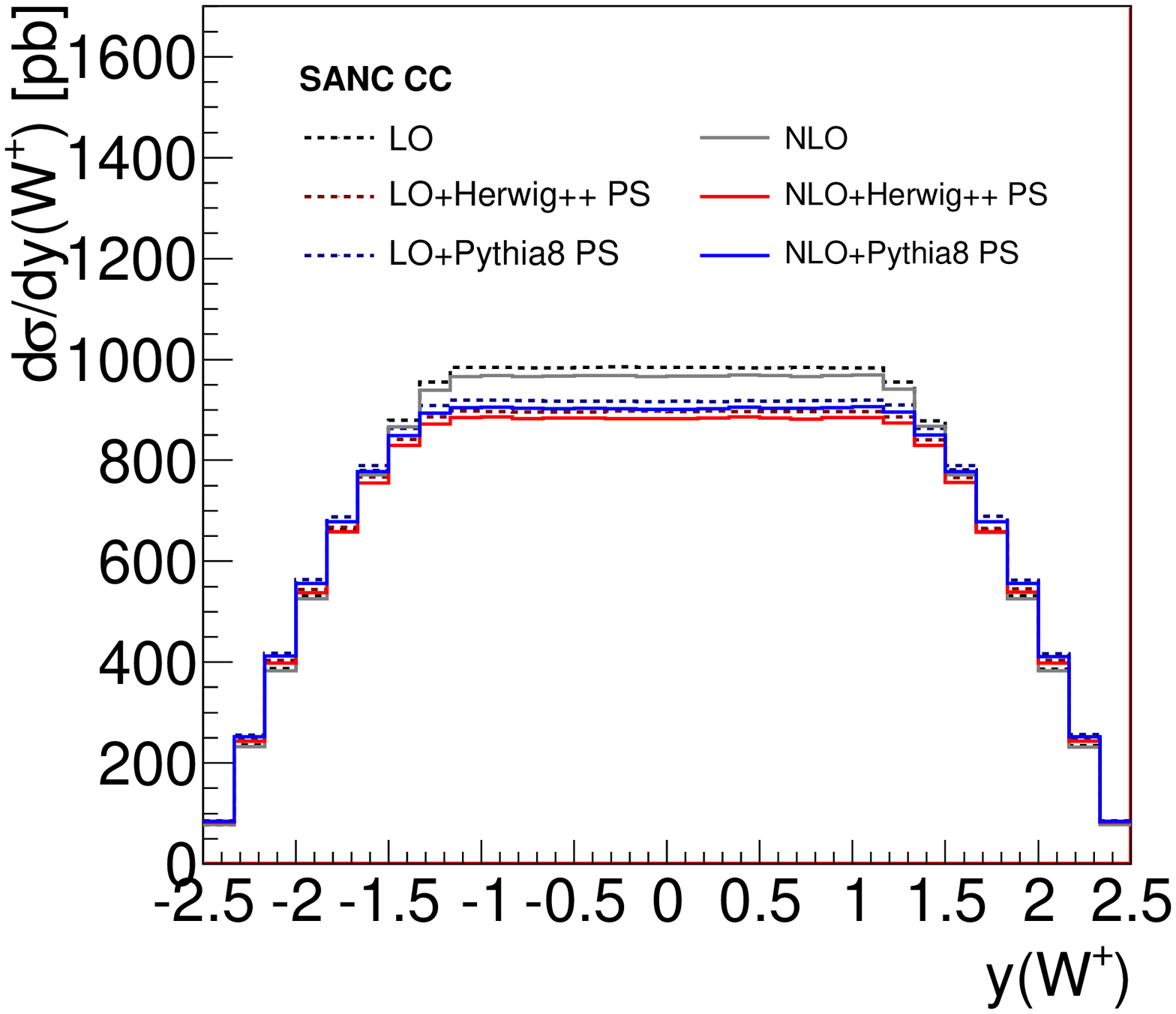}
\includegraphics[width=6.0cm, height=5.0cm]{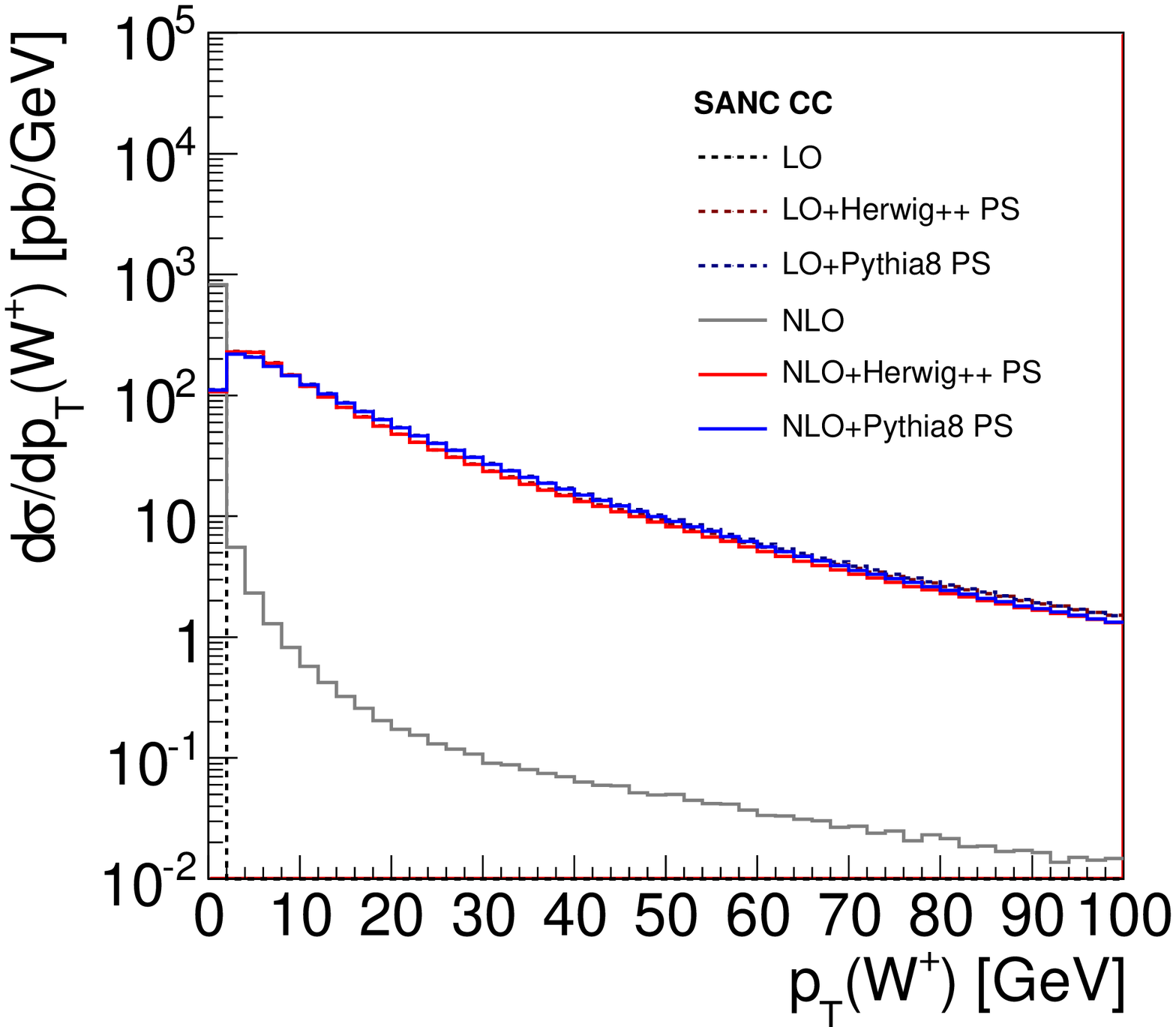}}\\
\subfloat{
\includegraphics[width=6.0cm, height=5.0cm]{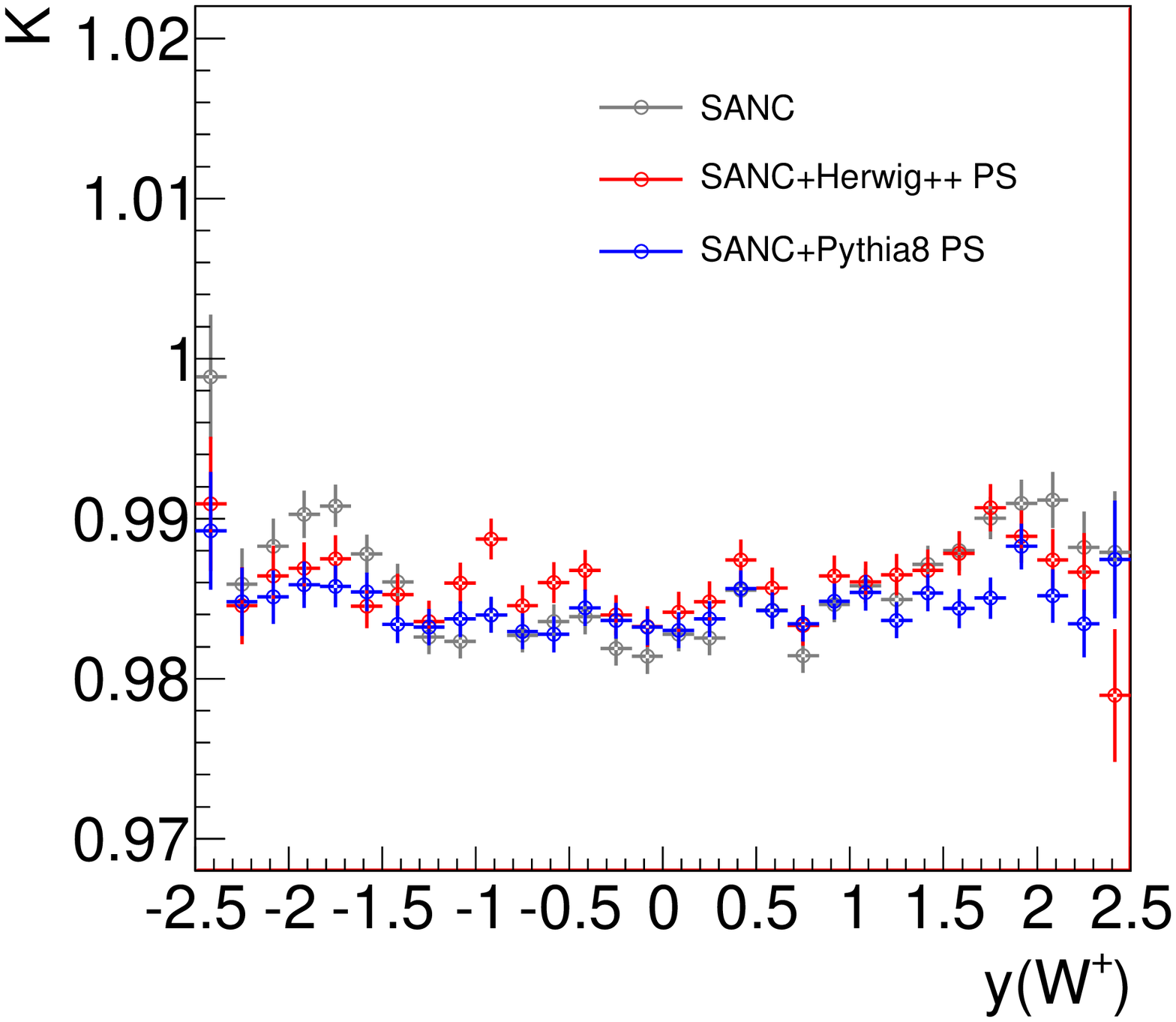}
\includegraphics[width=6.0cm, height=5.0cm]{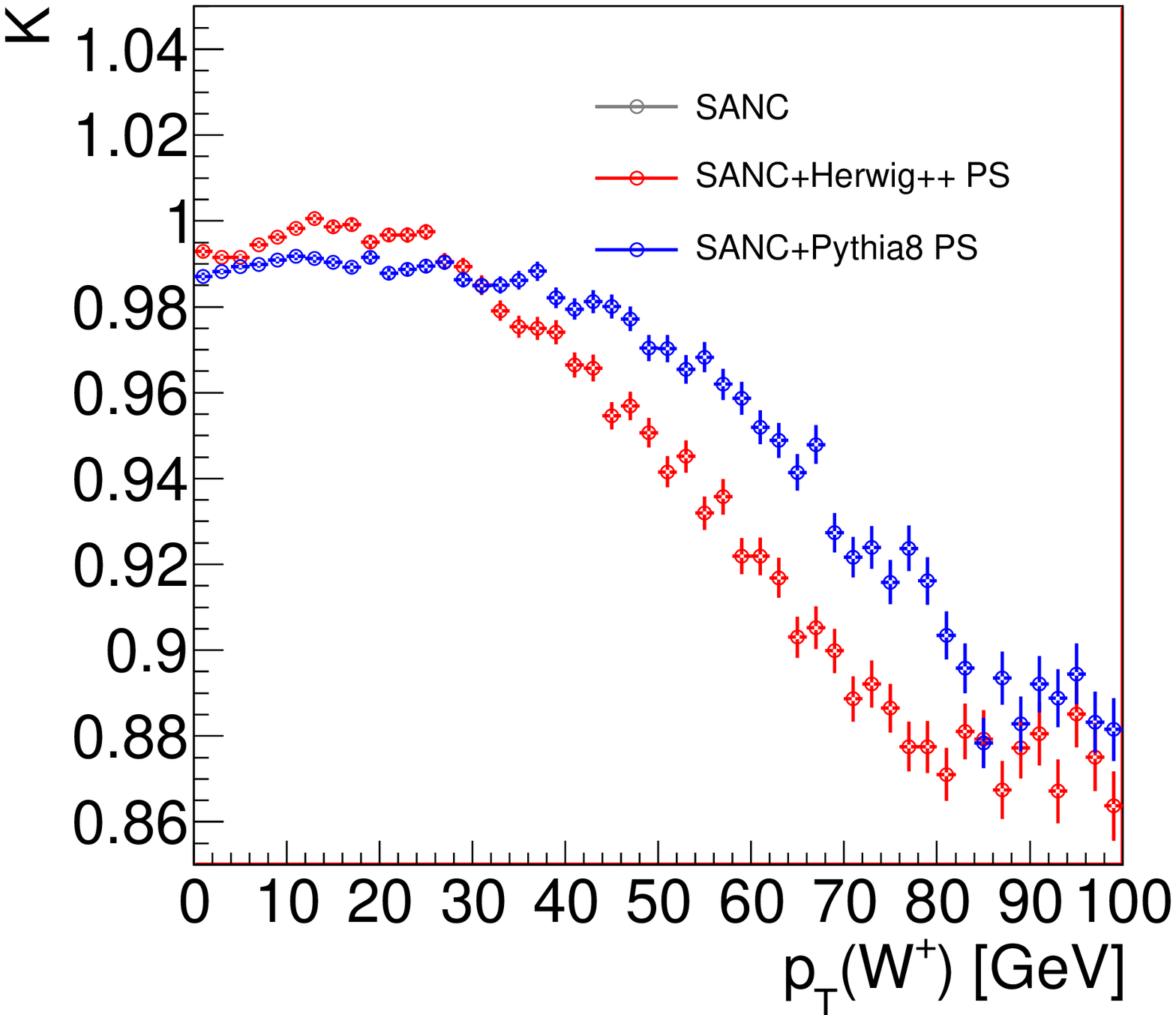}}\\
\subfloat{
\includegraphics[width=6.0cm, height=5.0cm]{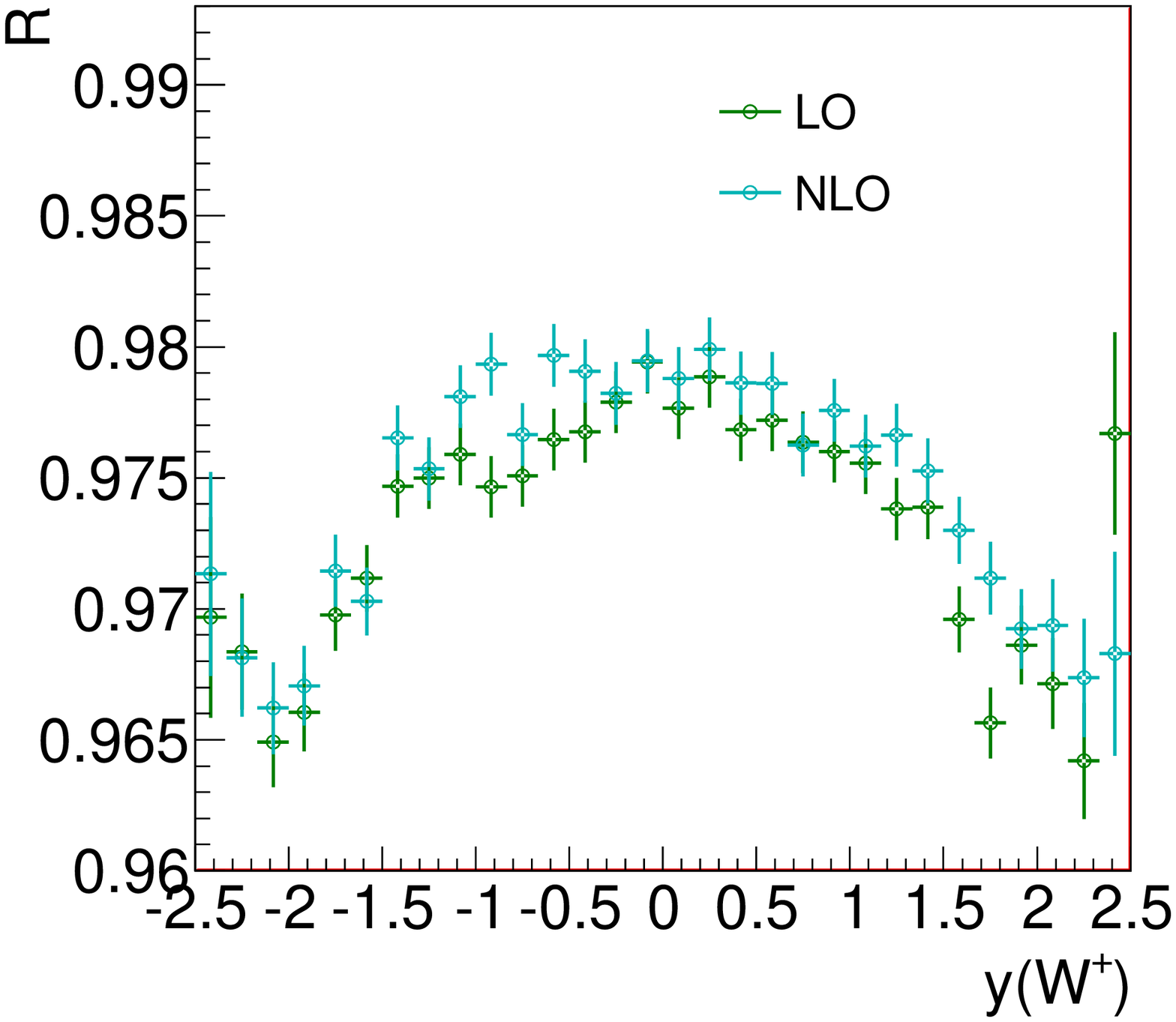}
\includegraphics[width=6.0cm, height=5.0cm]{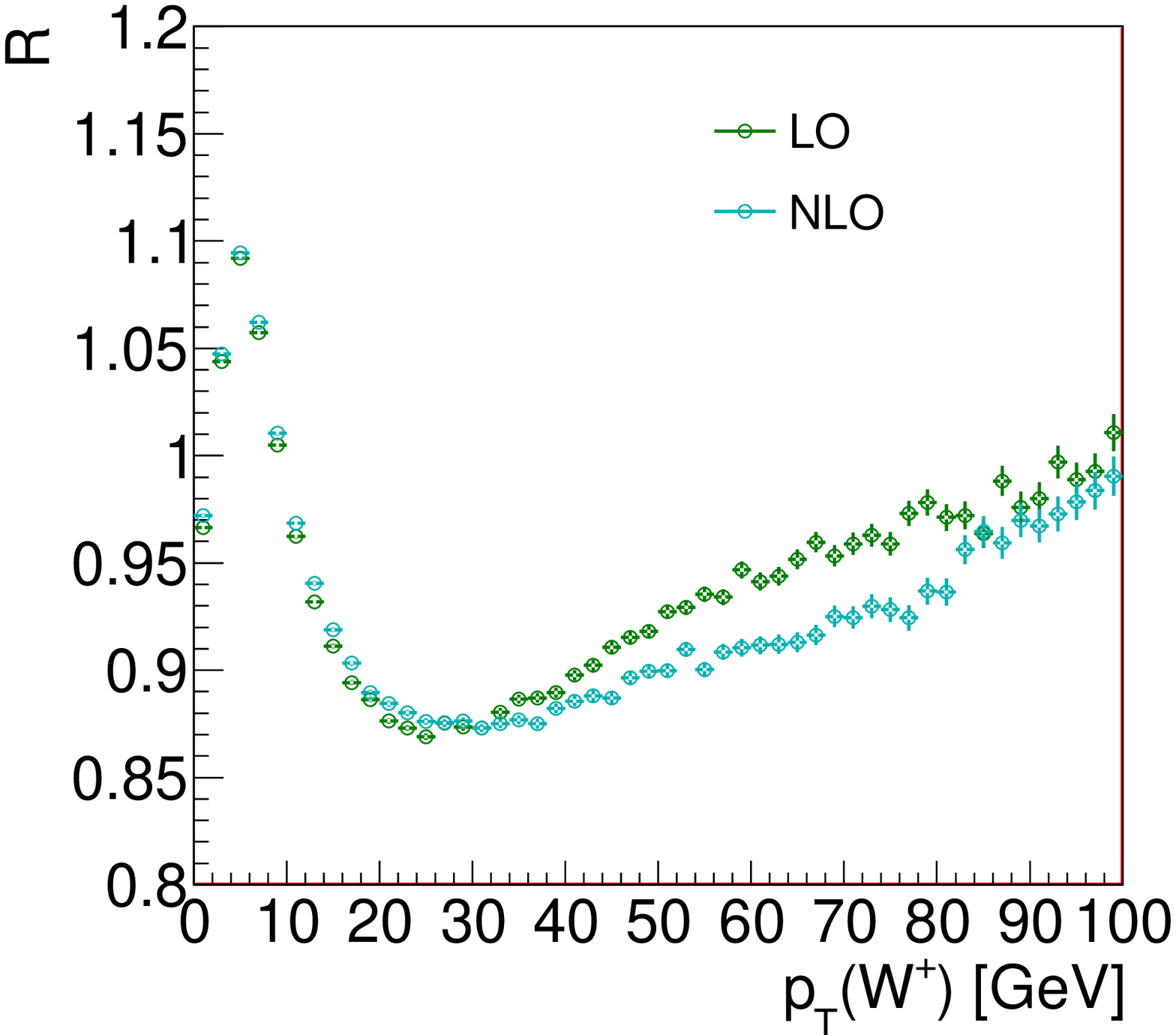}}
\end{center}
\caption {Distributions of rapidity (\textit{left}) and transverse momentum
(\textit{right}) of $W^+$ boson for CC DY.}
\label{fig_cc_W}
\end{figure}
\begin{figure}
\begin{center}
\includegraphics[width=6.0cm, height=5.0cm]{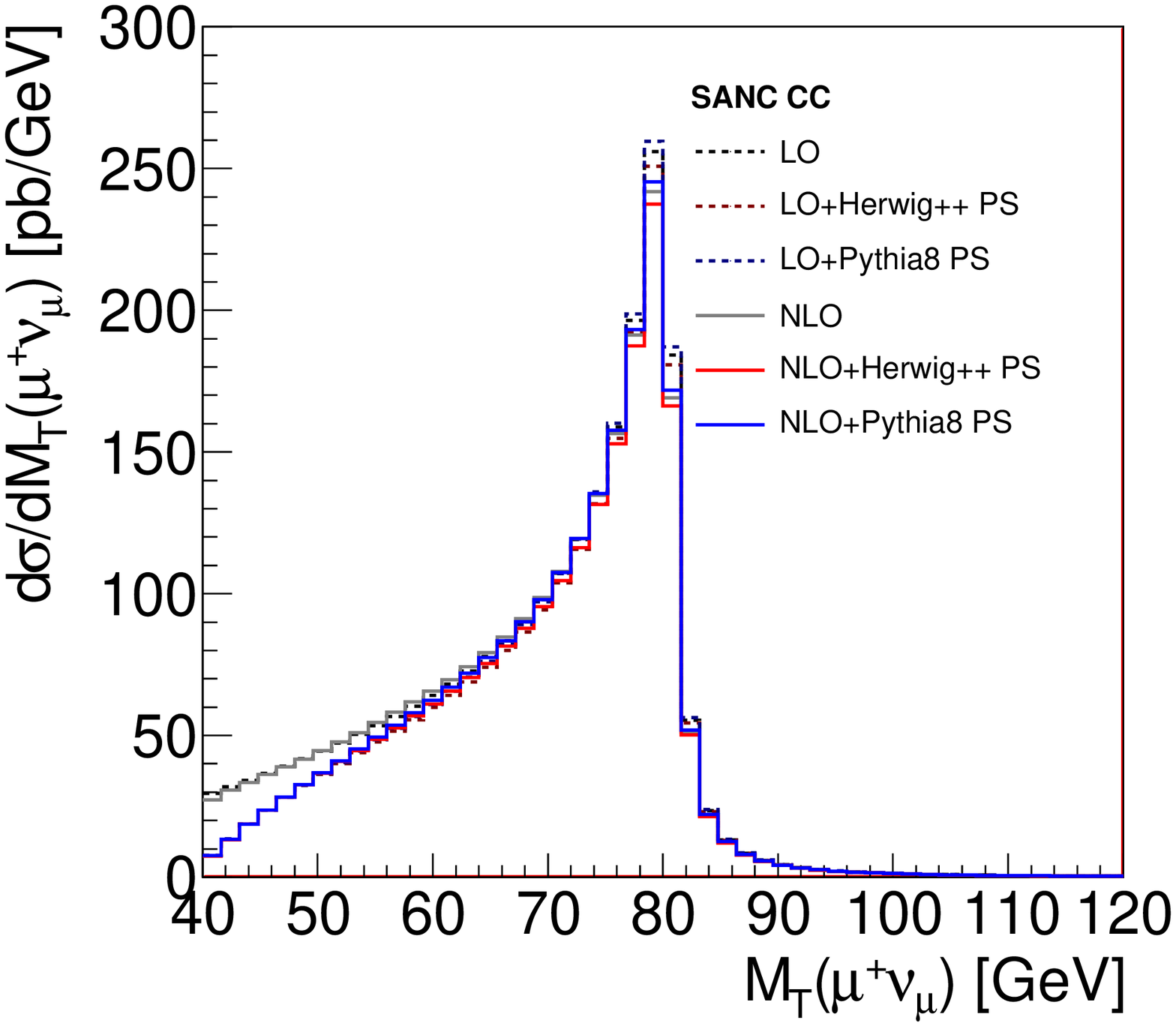}\\
\includegraphics[width=6.0cm, height=5.0cm]{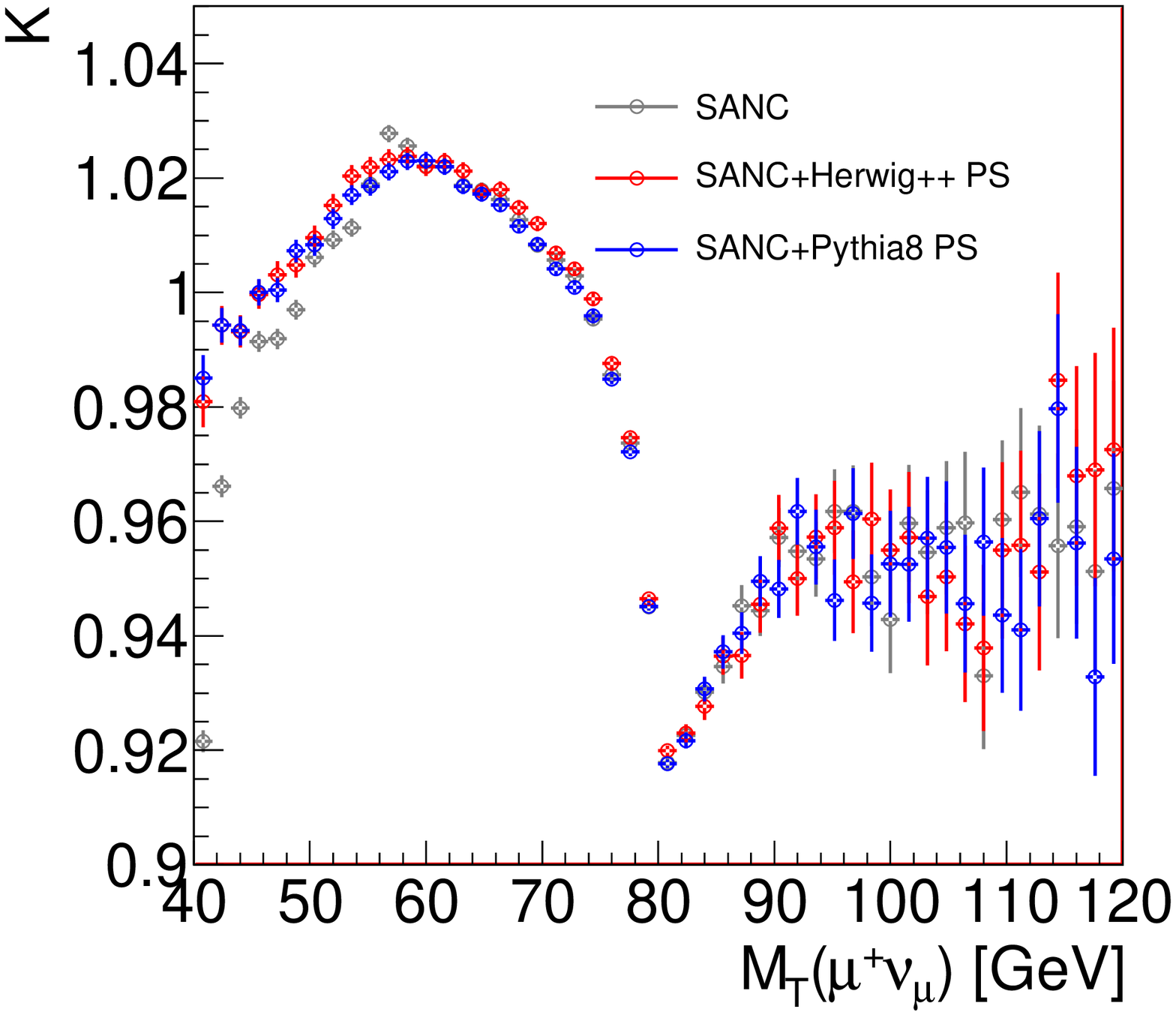}\\
\includegraphics[width=6.0cm, height=5.0cm]{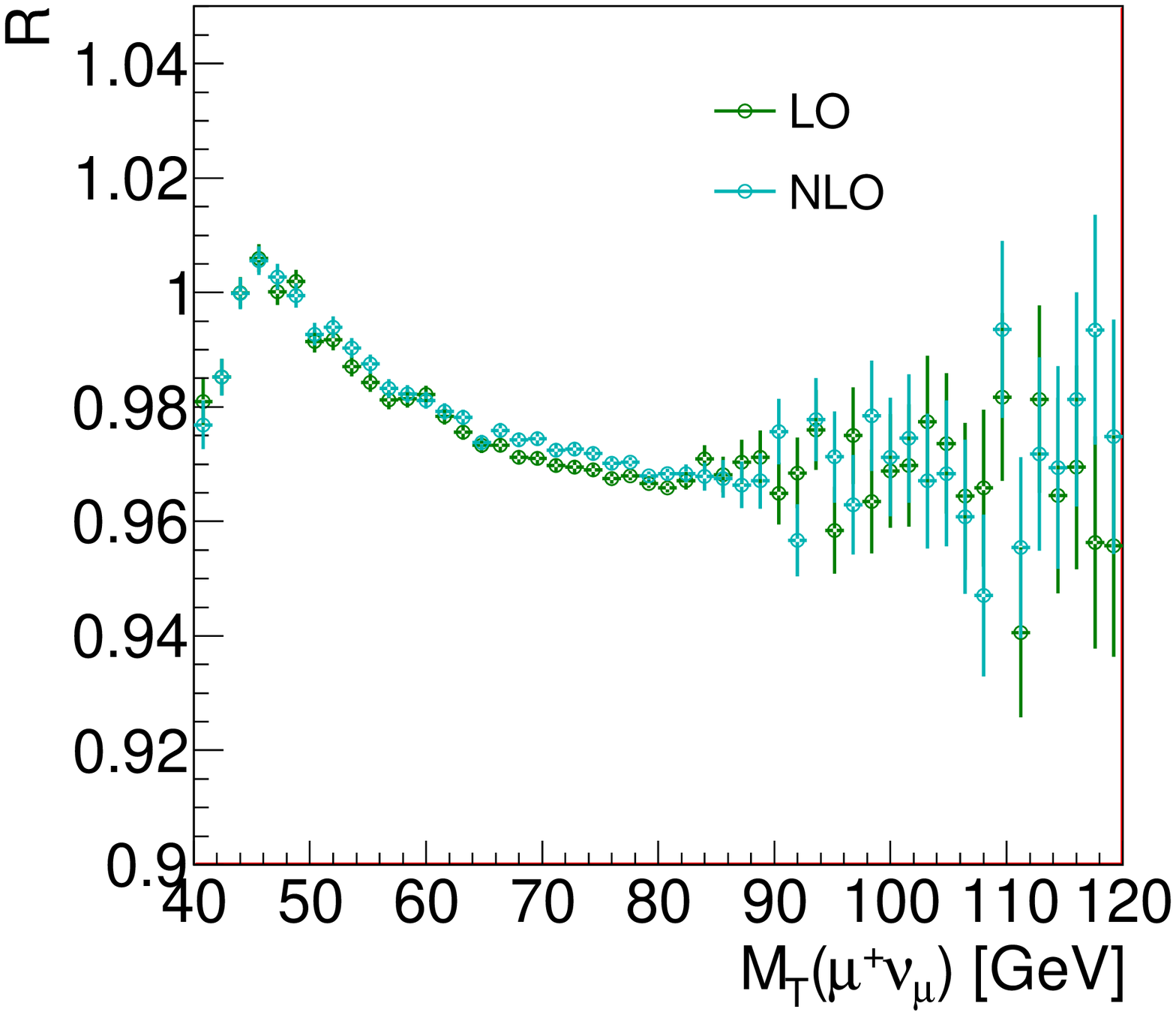}
\end{center}
\caption {Distributions of transverse mass of $\mu^+, \nu_\mu$ pair for CC DY.}
\label{fig_cc_mperpW}
\end{figure}
Figures \ref{fig_nc_mu}--\ref{fig_cc_mperpW} show distributions for various
observables obtained after the $cut 2$ selection. Each figure contains three
rows of plots with distribution of the differential cross sections
themselves (\textit{top} row), electroweak $K$-factor which is defined as
usual as $K = \sigma_{NLO}/\sigma_{LO}$ (\textit{middle} row) and the $R_X$
value (\textit{bottom} row).  The distributions show that $R_X$ values can
differ from unity by up to $10\%$ for $p_{T}$ (right columns in
figures~\ref{fig_nc_mu},~\ref{fig_nc_Z},~\ref{fig_cc_mu},~\ref{fig_cc_W}) and by
several percent for other observables.

The difference between the shower algorithms is most noticeable in the $p_{T}$
distributions. Nevertheless $R_X$
distributions in $M_{inv}(\mu^+ \mu^-)$ and $M_{T}(\mu^+ \nu_{\mu})$ are
practically flat and differ from unity by only 2--3\%.

It should here be emphasized that the prescription presented in
this paper only concerns incorporating the first order of EW
corrections into a shower framework. In particular, the description of QCD
corrections is still handled only with leading-logarithmic precision,
and does not include any matching to higher-order QCD matrix elements
(see, e.g., \cite{Lenzi:2009fi,Buckley:2011ms}). Thus, the description of vector boson plus
jets can only be expected to be correct
for jets with $p_T\ll m_Z$ (representing the bulk of the cross
section). For harder jets, differences between Herwig++ and Pythia8
reflect the uncertainty associated with QCD corrections beyond LL. 
Further work would be required to include QCD matrix-element
corrections in this region.

The right columns of the plots in figures~\ref{fig_nc_mu},~\ref{fig_cc_mu} show
the muon transverse momentum in the $pp\to \mu^+ \nu_{\mu}+X$ and $pp\to \mu^+
\mu^- +X$ processes.  Although the radiative corrections are washed out by
parton showers in the peak region they reach up to $15\%$ for higher $p_T$
values.  The difference in the parton shower algorithms for EW RC is mildly
noticeable at $p_{T}>50$~GeV where the $K$-factors for Pythia8 and Herwig++
diverge with maximum $2\%$.  The small bends in the $20$~GeV region are edge
effects that appear in the showered events selection and play no role in the
physics of the process.

A similar behaviour can be seen in the $Z/W$ transverse momentum distributions
in figures~\ref{fig_nc_Z},~\ref{fig_cc_W}: the $K$-factors deviate up to
$4\%$. The $(\mu^+ \mu^-)$ invariant mass and $(\mu^+ \nu_{\mu})$ transverse
mass plots in figures~\ref{fig_nc_minvZ},~\ref{fig_cc_mperpW} show no
significant effects.

\section{Summary \label{sum_sect}}

An interface between the SANC matrix-element generator and the Herwig++ and
Pythia8 parton shower Monte Carlo codes has been presented. As part of this
work, the new possibility of backwards evolution of photons has been added to
both the Herwig++ and Pythia8 initial-state showers.  Several numerical
crosschecks have been performed, with reasonable results. The addition of
parton showering gives a natural smearing effect on the EW K-factor
distributions in the Drell-Yan process. The lepton \(p_T\) distributions are
mostly affected in the Z and W peak region.

The remaining difference between Pythia8 and Herwig++ showering algorithms was
another focus of this study. The comparative plots included show that the
difference in differential cross section can reach up to 10\% for certain
observables. We expect that this could be further reduced by extending
the prescription presented here to include a matching to fixed-order QCD
matrix elements for vector boson plus jets.   

Since the completion of this work, two implementations of electroweak
corrections to W boson production in the POWHEG framework have appeared
\cite{Bernaciak:2012hj, Barze:2012tt}, combining both EW and QCD corrections.
However these works do not take into account effects of photon-induced
processes. We consider that the implementation into a general-purpose
electroweak tool like SANC has advantages for precision EW studies, given the
importance of EW scheme-dependence in radiative corrections and the need for
consistent scheme implementations between different processes.  Nevertheless,
it is clear that the POWHEG framework is an extremely powerful tool in
describing and combining hard QCD and parton shower corrections consistently,
and we look forward to making detailed comparisons between the results of these
implementations and our own.
 
\acknowledgments{This work was supported in part by the Marie
Curie research training network ``MCnet'' (contract number MRTN-CT-2006-035606),
by the RFBR grants 07-02-00932, 10-02-01030 and by the Dynasty Foundation.}

\bibliographystyle{JHEP}
\bibliography{EWRC_PS}

\end{document}